\documentclass[a4paper,12pt]{article}

\pdfoutput=1 

\usepackage[hmargin=.7in,vmargin=1.1in]{geometry}
\usepackage{indentfirst}
\usepackage{amsfonts}
\usepackage{mathrsfs}
\usepackage{amsmath}
\usepackage{amssymb}
\usepackage{authblk}
\usepackage{cite}
\usepackage[table]{xcolor}
\usepackage{xcolor}
\usepackage{mathtools}
\usepackage{graphicx}

\usepackage{bm}
\usepackage{upgreek}
\usepackage{braket}
\usepackage{color,soul}
\usepackage{csquotes}
\usepackage{caption}
\usepackage{subcaption}
\usepackage{orcidlink}
\usepackage{verbatim}
\usepackage{booktabs}
\providecommand{\colrule}{\midrule}
\providecommand{\botrule}{\bottomrule}

\hypersetup{
            colorlinks,
            linkcolor=[rgb]{0,0.3,0.6}, 
            citecolor=[rgb]{0,0.3,0.6}, 
            urlcolor=[rgb]{0,0.3,0.6}
           }

\def\be{\begin{eqnarray}}
\def\ee{\end{eqnarray}}

\begin{document}

\title{\Large\textbf{Quasinormal modes of a scale-dependent black hole in 5D spacetime with cosmological constant}}


\author[a,b]{
\'Angel Rinc\'on {\orcidlink{0000-0001-8069-9162}} \thanks{e-mail: 
\href{mailto:angel.rincon@physics.slu.cz}
{\nolinkurl{angel.rincon@physics.slu.cz}}}
}

\author[b]{
Zdeněk Stuchlík  \orcidlink{}
\thanks{e-mail: 
\href{mailto:zdenek.stuchlik@physics.slu.cz}
{\nolinkurl{zdenek.stuchlik@physics.slu.cz}}}
}


\affil[a]{
\normalsize{\em Departamento de F{\'i}sica, Universidad del B{\'i}o-B{\'i}o, Casilla 5-C, Concepci{\'o}n, Chile.}
}

\affil[b]{
\normalsize{\em Research Centre for Theoretical Physics and Astrophysics, Institute of Physics, Silesian University in Opava, Bezrucovo nám. 13, 74601 Opava, Czech Republic.}
}

\date{ }

\maketitle

\begin{abstract}
In this paper, we investigate the quasinormal modes (QNMs) of a scale-dependent black hole in five dimensions with a positive cosmological constant, considering i) scalar, ii) electromagnetic, and iii) Dirac perturbations.
After deriving exact analytical solutions within the framework of scale-dependent gravity in five dimensions and selecting a different family of solutions (cf. \cite{Contreras:2019fwu}), we analyze the corresponding effective potentials (given by the lapse function) for each perturbation type. 
We employ a first-order approximation for the lapse function, as it closely matches the exact case for small $\varepsilon$. Since quantum corrections are expected to be small, our focus is on a relatively low range of the scale-dependent parameter $\varepsilon$.
We confirm that the WKB semi-analytical approach can be used to compute the QNMs for these perturbations to the background geometry.
We calculate the QNM frequencies for the classical case (for comparison) and for cases where the scale-dependent parameter $\varepsilon$, which quantifies deviations from General Relativity, is non-zero. The results are summarized in a series of tables and figures.
Finally, we find that, regardless of the perturbation type, the solutions are stable, as the imaginary part of each QNM frequency is negative.
\end{abstract}


\tableofcontents

\newpage
\section{Introduction}
\label{sec:intr}

The pursuit of a quantum theory of gravity remains one of the most profound challenges in theoretical physics. Among the promising frameworks (like loop quantum gravity \cite{Ashtekar:2021kfp} or Causal dynamical triangulation \cite{Loll:2019rdj}), asymptotically safe gravity \cite{Eichhorn:2020mte} stands out by proposing that gravity achieves non-perturbative renormalizability through an ultraviolet fixed point \cite{Reuter:1996cp}. 
This paradigm enables the inclusion of quantum corrections into classical gravitational systems, particularly in extreme curvature regimes such as those found in black holes.

Knowing that the simplest solutions of Einstein’s field equations in general relativity (and in many alternative theories of gravity) are black holes, their study is essential for deepening our understanding of the fundamental laws of nature \cite{Stephani:2003tm}. As extreme astrophysical objects, black holes probe the very limits of both general relativity and quantum mechanics, thereby offering a unique setting to investigate their interplay. 
For example, black hole thermodynamics is intrinsically governed by the presence of the event horizon \cite{Witten:2024upt,Carlip:2008wv,Bardeen:1973gs}. Furthermore, black holes provide an ideal arena for the predictions of modified theories of gravity, enabling the exploration of potential deviations from GR. 
From the quantum perspective, the discovery of Hawking radiation (an remarkable quantum effect) demonstrated that black holes emit energy in a manner analogous to blackbody radiation ~\cite{Hawking:1974rv,Hawking:1975vcx}. 
In parallel, Bekenstein proposed that black holes possess an entropy, $S$, proportional to the area of their event horizon, $A_H$ ~\cite{Bekenstein:1973ur,Bekenstein:1974jk}. These seminal results suggested that black holes are not solely geometric configurations but also thermodynamic systems. 
Thus, a comprehensive understanding of black hole physics necessarily requires the synthesis of three fundamental pillars: (i) thermodynamics, (ii) gravity, and (iii) quantum physics~\cite{Wald:1999vt,Carlip:2014pma,Carlip:2001wq,Nozari:2006ka}. 
Of special interest is the cosmological constant due to its role in astrophysical processes \cite{Stuchlik:2020rls,Stuchlik:2025fws}.
However, combining these ingredients remains highly non-trivial, as their mutual interplay introduces profound and often unexpected effects in black hole physics. 

From the various approaches where quantum effects are included and applied to the context of black hole physics, it is essential to point out the renormalization group improvement of black hole spacetimes by Bonanno and Reuter \cite{Bonanno:2000ep}. Their work demonstrated how scale-dependent couplings, derived from renormalization group flows, can mitigate singularities and alter the thermodynamic properties of classical Schwarzschild black holes.
This contribution gave rise to a new research line, producing a variety of results (see, for example, 
\cite{Ishibashi:2021kmf,Koch:2014cqa} and the references therein).
Inspired by asymptotically safe gravity, the scale-dependent gravity formalism provides an alternative and simpler approach to include quantum effects by allowing classical constants (such as Newton's gravitational constant and the cosmological constant) to vary with a spacetime-dependent scale \cite{Koch:2016uso,Rincon:2020iwy,Contreras:2017eza,Contreras:2018dhs,Rincon:2018lyd,Rincon:2019zxk,Fathi:2019jid,Panotopoulos:2021tkk,Bargueno:2021nuc,Ovgun:2023ego,Rincon:2023kun,Calzada:2023yiq,Koch:2025gaw,Contreras:2018gpl,Contreras:2018gct}. 
In spherically symmetric spacetimes, this auxiliary scale typically corresponds to the radial coordinate only, facilitating the derivation of modified black hole solutions that approximate quantum corrections.
A vast part of the literature is devoted to four-dimensional scenarios, however, higher-dimensional gravity, motivated by string theory and braneworld models, offers a richer context for exploring black hole physics \cite{Emparan:2008eg}. In five dimensions, neutral black holes in a de Sitter background, with a positive cosmological constant, exhibit unique properties relevant to cosmological models and analogs of the AdS/CFT correspondence \cite{Birmingham:1998nr}.

Of the many remarkable properties of black holes, their stability under various perturbations, known as quasi-normal modes (QNMs), is particularly notable \cite{Berti:2009kk,Nollert:1999ji,Konoplya:2011qq}. This matter has attracted considerable interest since the first detection of gravitational waves a decade ago \cite{LIGOScientific:2016aoc}. Although QNMs studies have been explored in different spacetimes, there are still many theoretical frameworks that offer opportunities to further investigate these characteristic oscillations.

In this study, we re-derive and analyze a neutral black hole solution in five-dimensional scale-dependent gravity with a positive cosmological constant focusing in a family of solutions alternative (slightly different) of our previous work \cite{Rincon:2020iwy}). 
To investigate the stability and dynamical response of this solution, we compute the quasinormal modes for scalar, electromagnetic, and Dirac massless perturbations using the Wentzel-Kramers-Brillouin (WKB) approximation.
In addition, the WKB method (refined to higher orders) provides reliable computations for the dominant QNM frequencies, which describe the damped oscillations during the black hole ringdown phase \cite{Schutz:1985km, Iyer:1986np, Konoplya:2019hlu}. These modes are essential for evaluating black hole stability and have implications for gravitational wave astronomy.
Our results highlight the impact of the scale-dependent parameter on the QNM spectra, revealing subtle quantum modifications to classical behaviors. The consistently negative imaginary parts of the frequencies confirm stability across all perturbation types.

The paper is structured as follows: following this concise introduction, Section \eqref{sec:moti} reviews the motivation and formalism of scale-dependent gravity. In Section \eqref{sec:back}, we derive the black hole solution and identify the appropriate parameter to recover the classical background in the relevant limit. Section \eqref{sec:pert} then outlines the perturbations studied in this work, with a focus on their effective potentials. The WKB method is introduced in Section \eqref{sec:meth}, followed by Section \eqref{sec:nume}, which presents the numerical results, summarized through figures and tables. Finally, Section \eqref{sec:conclu} provides a brief summary of our key findings.

\section{Motivation and Formalism}
\label{sec:moti}

GR is considered an incomplete theory (not the final one) because
of allowing physical singularities, among other things, it cannot be quantized (at least up to know) given that it is a non-renormalizable theory \cite{Shomer:2007vq}. 
However, if we treat GR as an effective theory, we can incorporate quantum features with minimal modifications. The starting point is the core object responsible to generate a self-consistent theory, the so-called "effective action". Inspired by the classical action, where the coupling constant are taken to constants values, we extend the object to describe a gravitational theory in which quantum features could be consistently included. 
Roughly speaking, The effective action is a key concept in theoretical physics (particularly in quantum field theory and gravitational theories), that encapsulates the effects of quantum fluctuations on a system. As was previously said, it object replaces the classical action, and it is able to incorporating quantum corrections, providing a framework to obtain effective equations of motion. The effective action has convenient and well-defined properties, for example:  
  i) the effective action incorporates quantum effects,
 ii) the effective action is reduced to the classical action when the quantum correction vanishing, 
iii) the effective action accounts for renormalization effects, leading to running coupling constants.
The later features is precisely exploited in a variety of formalism based on asymptotically safe gravity, being one of them the so-called scale-dependent gravity.
In such a formalism, the gravitational effective action, $\Gamma[k, g_{\mu \nu}, \cdots]$, where $k$ is a scale field, and its describes the
theory. 
Becomes essential to mention that the Asymptotic Safety program (which is based on a non-trivial ultra-violet fixed point for the leading dimensionless gravitational couplings) is the conventional scenario where these ideas have been satisfactorily used. In addition, should be highlighted that was Weinberg, in his seminal work \cite{Weinberg:1976xy}, who introduced this program. Substantial improvement has been made up to now \cite{Wetterich:1992yh,Morris:1993qb,Reuter:1996cp,Reuter:2001ag,Litim:2002xm,Litim:2003vp,Niedermaier:2006wt,Niedermaier:2006ns,Gies:2006wv,Machado:2007ea,Percacci:2007sz,Codello:2008vh,Benedetti:2009rx,Manrique:2009uh,Manrique:2010am,Manrique:2010mq,Eichhorn:2010tb,Litim:2011cp,Falls:2013bv,Dona:2013qba,Falls:2014tra,Eichhorn:2018yfc,Eichhorn:2017egq}.

Ignoring the details, there are three main forms to account quantum corrections into classical gravitational solutions \cite{Reuter:2003ca,Rincon:2022hpy}: 
  i) at the level of the solution,
 ii) at the level of the equations of motion, and
iii) at the level of the action.
As we mentioned early, a notable example is the seminal work of Bonanno and Reuter \cite{Bonanno:2000ep}. To be more precise, their research thoroughly explores how renormalization group effects alter the classical Schwarzschild black hole solution, providing new insights into its properties. 
Such scenario does not yield exact solutions to the effective quantum-field equations and thus cannot fully capture the quantum nature of spacetime. However, it offers essential insights into the primary effects distinguishing a classical black hole from its quantum counterpart.
In scale-dependent gravity, the central object $\Gamma[k, g_{\mu \nu}, \dots]$ accounts quantum modifications by allowing $k \equiv k(x^{\mu})$ to encapsulate quantum deviations. Consequently, the structure of the theory is modified such that the classical couplings in the action, such as Newton's constant $ G_0 $ or the cosmological constant $ \Lambda_0 $, become functions that vary with a specific scale, typically the radial coordinate in the simplest case.
In the scale-dependent formalism, we derive the effective field equations by varying the effective action with respect to:
i) the metric field and
ii) the auxiliary field $ k $.

While it is possible to solve these equations, finding exact solutions under condition ii) is very challenging. To address this issue, we usually impose a practical constraint on the effective energy-momentum tensor to satisfy the null energy condition. Although this is an assumption, it is satisfied by many well-defined black hole solutions, and it is the least restrictive of all energy conditions, allowing for a broad range of analytical solutions. This constraint produces an extra differential equation for the Newton coupling, which closes the system of equations and enables us to derive exact solutions that correspond to the classical limit, in which the coupling functions become constants.

For a line element satisfying the Schwarzschild gauge, no additional constraints are required. We simply vary the effective action with respect to the metric to derive the relevant equations of motion.


\section{Background: Review and improvement}
\label{sec:back}

From now on, we will restrict our attention to the gravity in 5D spacetimes \footnote{A special kind of 5D spacetimes is representing by the braneworld  models \cite{Randall:1999vf,Randall:1999ee,Maartens:2010ar}.}.
As we mentioned before, in the scale-dependent formalism the classical couplings are promoted to functions which, taking advantage of the symmetry, depends on radial coordinate only, i.e., $G_N \rightarrow G(r)$ and $\Lambda_N \rightarrow \Lambda(r)$ for the varying Newton and cosmological couplings, respectively. 
To maintain the paper self-contained, we will start by including the average effective action, $\Gamma[g_{\mu\nu},k]$, solely with three ingredients: 
i) the Ricci scalar, $R$, 
ii) the cosmological function, $\Lambda_k$, and 
iii) the matter action, $\Gamma_{\text{M}}$, 
while, the classical Einstein coupling, $\kappa_0$, is replaced by its scale-dependent counterpart $\kappa_k\equiv8\pi G_k$, being $G_k$ the scale-dependent Newton's coupling. Then, the corresponding action is
\begin{equation}
\Gamma[g_{\mu\nu},k]=\int \mathrm{d}^5 x \sqrt{-g}\left[\frac{1}{2\kappa_k}\Bigl(R-2\Lambda_k\Bigl)\right] + \Gamma_{\text{M}},
\end{equation}
where the speed of light is taken to be one, while $\Lambda_k$ is the cosmological couplings respectively. 
Taking the corresponding variation with respect to the inverse metric field $g^{\mu \nu}$, we obtain the modified form of Einstein's field equations,  
\begin{equation}
E_{\mu \nu} \equiv G_{\mu\nu} + g_{\mu\nu}\Lambda_k + \Delta t_{\mu \nu} =0.   
\end{equation}
Note that an effective energy-momentum tensor can then be naturally defined and the effective Einstein field equations rewritten, i.e., 
\begin{equation}
G_{\mu\nu}+\Lambda_k g_{\mu\nu}=\kappa_k (T^{\textrm{eff}})_{\mu\nu},
\label{einstein}
\end{equation}
being $(T^{\textrm{eff}})_{\mu\nu}$ the corresponding effective energy-momentum tensor, defined according to
\begin{equation}
(T^{\textrm{eff}})_{\mu\nu}:=(T_{\mu\nu})_k-\frac{1}{\kappa_k}\Delta t_{\mu\nu},
\end{equation}
where the auxiliary tensor $\Delta t_{\mu\nu}$ arises from the scale dependence of Newton's constant and is defined as
\begin{equation}
\Delta t_{\mu\nu} = G_k \left( g_{\mu\nu}\,\square - \nabla_\mu \nabla_\nu \right) G_k^{-1}.
\end{equation}
Please, be aware and notice that as we are in a five dimensional background, it is not necessary to consider any supplementary condition to close the system of differential equations, as usually occurs in lower dimensional cases. Instead, given that we are considering spherical symmetry and taking advantage of the Schwarzschild ansatz, we have to determine an unique metric function plus two additional scale-dependent functions  $\{f(r), G(r), \Lambda(r) \}$, by solving the Einstein fields equations. Let us start by consider a line element in five dimension with spherical symmetry, taking the coordinates $x^{\mu}=\{t,r,\sigma, \theta, \phi \}$:
\begin{align}
\mathrm{d}s^2 &= -f(r) \mathrm{d}t^2 + f(r)^{-1} \mathrm{d}r^2  + r^2 \Bigl(\mathrm{d}\sigma^2 + \sin^2(\sigma) \mathrm{d}\Omega^2\Bigl),
\end{align}
where $\mathrm{d}\Omega^2 = \mathrm{d}\theta^2 + \sin^2(\theta)\mathrm{d}\phi^2$.
Firstly, in order to solve the functions easily, and as the cosmological function appears as an algebraic term only, we remove it from the effective equations. After do that, we use the simplified  $E_{00}$ equation and defining a new auxiliary
\begin{align}
      \beta(r) &\equiv \frac{r^2 G'(r)}{2 G(r)}, 
\end{align}
we get the non-linear differential equation for $\beta(r)$ in the form
\begin{align}
    -\frac{2 \beta (r)}{r^2}+\frac{\beta '(r)}{\beta (r)}-\frac{2}{r} = 0 ,
\end{align}
where $\beta(r)$ appears as consequence of the running of the gravitational coupling.
Solving last equation and expressing the solution in term of the Newton function, we obtain the well-know scale-dependent Newton coupling $G(r)$:
\begin{align}
    G(r) &= \frac{B}{A + r} .
\end{align}
As a mandatory remark, let us mention that the form of the Newton coupling obtained for the five-dimensional solutions is general, irrespectively of the number of dimensions. In particular, as it was previously mentioned in preliminary papers \cite{Koch:2016uso,Rincon:2020iwy,Contreras:2017eza}, a suitable choice of certain null vector and the use of the saturated version of the null energy condition induce the same second-order differential equation for $G(r)$ (see, for instance \cite{Rincon:2017ayr}).
The scale-dependent lapse function $f(r)$ is subsequently obtained from the equation $E_{33}$ after: i) removing $\Lambda(r)$ and ii) replacing the solution of the Newton coupling. The differential equation for $f(r)$ takes the form 
\begin{align} \label{diff_f}
   \frac{1}{2} r^2 f''(r)+\left(\frac{1}{2}r - \beta (r)\right) f'(r) + \bigg(  \frac{4 \beta (r)^2}{r^2}-2 \beta '(r) + \frac{6 \beta (r)}{r} - 2 \bigg)f(r) + 2 = 0 .
\end{align}
Solving directly equation \eqref{diff_f} we obtain the scale-dependent lapse function:
\begin{align}
    f(r) = 1 &- \frac{L}{2 A^3} - \frac{L}{4 A r^2} + \frac{L}{3 A^2 r} 
    - \bigg[\frac{2}{3A} - \frac{L}{A^4}\bigg] r
    + \bigg[\frac{5}{9 A^2} +  J + \bigg(\frac{2}{3A^2} - \frac{L}{A^5} \bigg) \ln \bigg( 1 + \frac{A}{r} \bigg) \bigg] r^2.
\end{align}
Finally, having obtained $G(r)$ and $f(r)$, it is trivial to obtain directly $\Lambda(r)$ by replacement on the original effective Einstein field equations. Using the algebraic equation for the cosmological coupling we have
\begin{align}
    \Lambda(r) &= \left(\frac{\beta (r)}{r^2}-\frac{3}{2 r}\right) f'(r) + 
    \left(\frac{6 \beta (r)}{r^3}-\frac{3}{r^2}\right) f(r)
    + \frac{3}{r^2}  ,
\end{align}
and thus we confirm that the cosmological coupling takes the form
\begin{align}
\begin{split}
    \Lambda(r) = \frac{1}{18 A^5 r^2 (A+r)^2} 
    \Bigg[
    &-4 A r^3 \left(72 A^5 J+10 A^3+45 L\right) - 18 A^2 r^2 \left(6 A^5 J-3 A^3+11 L\right)
    \\
    &-20 A^3 r^4 \left(9 A^2 J+5\right) - 24 A^3 L r+3 A^4 L
    -12 r^2 \left(2 A^3-3 L\right) \times
    \\
    &\times (A+r) (3 A+5 r) \ln \left( 1 + \frac{A}{r}\right)
    \Bigg].
\end{split}
\end{align}

Keep in mind that the four-dimensional version of this solution was discovered years ago by Koch and colleagues, marking the first solution identified within this framework \cite{Koch:2015nva}.
It is convenient to mention that the differential equations for $G(r)$ and $f(r)$ are of second order, which means we have $2 \times2$ integration constants. In particular, the set of values $\{A, B, L, J\}$ should be adjusted in such a way that satisfy two main conditions:
  i) in the classical limit (absence of quantum corrections), these constants should be the classical coupling constants, making the solution mimic the classic case, 
 ii) in the scale-dependent case (i.e., with non vanishing quantum corrections), we should have a control parameter to encode the deviation with respect to the classical solution. In addition, we will focus on a subclass of solutions in which the impact of the scale-dependent parameter on the cosmological coupling is reduced.
With the above in mind, we rewrite the integration constants in term of parameters of the classical solution, namely:
\begin{align}
    \frac{L}{4 A} &\equiv 2 M_0 G_0,
    \\
     J + \frac{5}{9 A^2} &\equiv - \frac{1}{6} \Lambda_0
\end{align}
Finally, the choice of the constants $A$ and $B$ is delicate: they must be chosen in such a way that i) it is possible to study both the classical regime and quantum effects,  ii) such a control parameter is desirable to be dimensionless. Taking into account the last two statements, one suitable choose is
\begin{align}
    A &\equiv \sqrt{G_0 M_0} \cdot \left(\frac{1}{\varepsilon} \right) ,
    \\
    B &\equiv A \cdot G_0.
\end{align}
At this point we have connected the integration constants with those which should appear in the classical solution $\{G_0, M_0, \Lambda_0\}$ plus one additional dimensionless constant $\varepsilon$ which encodes the quantum effects. Thus, when $\varepsilon$ is taken to be zero, the classic black hole solution should  be recovered. 
Once we replace the integration constants by the the classical coupling plus the scale-dependent parameter, the solutions are rewriting as follow: 
\\
i) the Newton function is finally adjusted to be
\begin{align}
    G(r) &= \frac{G_0}{1 + \varepsilon \cdot \left( \frac{r}{\sqrt{G_0M_0}} \right) },
\end{align}
ii) the lapse function takes the form:
\begin{align}
\begin{split}
    f(r) =  
    &- \frac{2 G_0 M_0}{r^2} + \frac{8 \sqrt{G_0 M_0} }{3 r} \ \varepsilon + (1 - 4 \ \varepsilon^2) - \Bigg( \frac{2 \ \varepsilon}{ 3 \sqrt{G_0 M_0} } - \frac{8 \ \varepsilon^3 }{\sqrt{G_0 M_0}} \Bigg) r 
    \\
    &- 
    \Bigg(
    \frac{1}{6}\Lambda_0 - \left(  \frac{2 \ \varepsilon^2}{3 G_0 M_0} - \frac{8 \ \varepsilon^4}{G_0 M_0}  \right) \ln \left( 1 + \frac{\sqrt{G_0 M_0}}{\varepsilon \ r}  \right)
    \Bigg)
    r^2.
    \end{split}
\end{align}
iii) the running cosmological coupling, after performing all the changes, is reduced to:
\begin{align}
\begin{split}
    \Lambda(r) = \ &\frac{1}{3 G_0 M_0 r^2 (\sqrt{G_0 M_0} + \varepsilon \ r)^2} 
    \Bigg[
    20 \sqrt{G_0 M_0} (1 - 12 \ \varepsilon ^2) \ \varepsilon^3 r^3 + G_0^2M_0^2 ( 4 \ \varepsilon^2 + 3 \Lambda_0 r^2 ) + 
    \\
    &G_0 M_0 \ \varepsilon r 
        \Bigl( 
        19 \ \varepsilon r - 32 \sqrt{G_0 M_0} \ \varepsilon^2 - 264 r \ \varepsilon^3 + 8 \sqrt{G_0 M_0} \Lambda_0 r^2 + 5 \Lambda_0 \ \varepsilon r^3
        \Bigl)
        - 
        \\
        &4 r^2 \ \varepsilon^2 (1 - 12 \ \varepsilon^2) 
            \Bigl[
            3G_0 M_0 + \varepsilon r (8 \sqrt{G_0 M_0} + 5 \varepsilon r)
            \Bigl] 
            \ln \left( 1 + \frac{\sqrt{G_0 M_0}}{\varepsilon \ r} \right)
    \Bigg].
    \end{split}
\end{align}
We check that when the scale-dependent effect is weak, the classical solution is slightly perturbed and, thus, when $\varepsilon \rightarrow 0$, the classical solution is completely recovered, as it should be by construction. Thus, for weak scale-dependent coupling
\begin{align}
    G(r) & = G_0 \Bigg(1 -  \varepsilon \cdot \frac{r}{\sqrt{G_0 M_0}} \Bigg) + \mathcal{O}(\varepsilon^2),
    \\
    f(r) &= f_0(r) +  \varepsilon \cdot \Bigg( \frac{8 \sqrt{G_0 M_0}}{3 r} - \frac{2r}{3 \sqrt{G_0 M_0}}  \Bigg) + \mathcal{O}(\varepsilon^2),
    \\
    \Lambda(r) &= \Lambda_0 \Bigg( 1 + \varepsilon \cdot \frac{2r}{3 \sqrt{G_0 M_0}}  \Bigg) + \mathcal{O}(\varepsilon^2).
\end{align}
and in the limit case, the three classic solutions are recovered, i.e., 
\begin{align}
    \lim_{\varepsilon \rightarrow 0} G(r) & = G_0 ,
    \\
    \lim_{\varepsilon \rightarrow 0} f(r) &\equiv f_0(r) =  1 - \frac{2 G_0 M_0}{r^2} - \frac{1}{6} \Lambda_0 r^2,
    \\
    \lim_{\varepsilon \rightarrow 0} \Lambda(r) &= \Lambda_0.
\end{align}
At this point, we should note that the proportionality constant of the cosmological term depends on the spacetime dimension. In this case, unlike the four-dimensional counterpart, the coefficient is $1/6$ instead of $1/3$ (in the classical limit).
In what follows, we will focus on positive values of the classical cosmological coupling, however, should be mentioned that, still in the classical case, there are a restriction for $\Lambda_0$ in terms of the product $G_0 M_0$. Thus, to obtain positively defined radius, we demand a positive discriminant $(D > 0)$ coming from $f_0(r)$. The algebraic equation is 
\begin{align}
        r^4 - \frac{6}{\Lambda_0}r^2 + \frac{12 G_0 M_0}{\Lambda_0} = 0,
\end{align}
Then, the condition for the discriminant induces the following restriction:
\begin{align}
    \Lambda_0 < \frac{3}{4 G_0 M_0} \equiv \Lambda_{\text{crit}}.
\end{align}
When this bound is satisfied, $f_0(r)=0$ admits two positive roots: the event horizon $r_H$ and the outer cosmological-like horizon $r_c$, with $r_H<r_c$. The static domain of outer communications is the interval $r_H<r<r_c$, in which the Killing vector $\partial/\partial t$ is timelike. For the small values of $\varepsilon$ considered below, the same two-horizon structure is preserved.
For the modified lapse function $f(r)$, the application of the same idea becomes complex for the inclusion of the logarithm function. 
To explore how quantum effects influence the cut-off for the cosmological coupling, we use the first-order approximation for the lapse function and calculate the discriminant based on the lower-order terms for $\varepsilon$.
Depending on the sign of $\varepsilon$, the expansion varies slightly. Thus, we can write the expression more compactly by accounting for the sign change in the third term. Specifically, when $\varepsilon$ is positive, we adopt the positive sign ($+$), whereas when $\varepsilon$ is negative, we take the negative sign ($-$). 
Roughly speaking, we can write the condition in the following form
\begin{align}
    \Lambda_0 \lesssim \Lambda_{\text{crit}}
    \bigg(
    1 + 
    \frac{16}{9}\varepsilon^2 
    \
    \pm \
    \frac{64}{27}\varepsilon^3 
    +
     \frac{64}{27}\varepsilon^4 + \cdots
    \bigg) \equiv \Lambda_{\text{crit}}(\varepsilon).
\end{align}
Thus, the quantum corrections induce a shift into the values that cosmological constant could take, feature which is absent in the classical case and parameterized in the last expression. For small values of $\varepsilon$, the allowed cosmological coupling is slightly larger than in the classical counterpart.

\begin{figure*}
    \centering
    \includegraphics[width=0.75\textwidth]{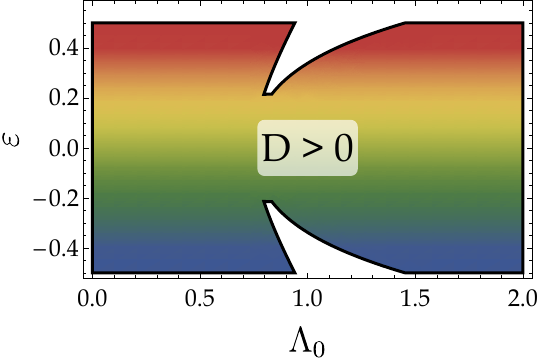}
    \caption{
Region plot of the discriminant for the first-order lapse function of the 5D scale-dependent black hole solution with positive cosmological constant.
    }
    \label{Fig:Disc}
\end{figure*}

For a qualitative comprehension, we show, in Figure \eqref{Fig:Disc}, when the discriminant is positive, for a concrete range of $\varepsilon$ and $\Lambda_0$.

\section{Perturbations}
\label{sec:pert}

In this section we will briefly summarize the main equations and ingredients required to understand three different type of perturbations: i) scalar, ii) electromagnetic, iii) Dirac fields, in the context of black hole physics.

\subsection{Scalar Perturbations}

Let us start by studying the dynamics of a test scalar field, labeled by $\Phi$, propagating in a 5D spacetime within a prescribed gravitational background. We also assume that the scalar field is real. Using the associated action, $S[g_{\mu \nu}, \Phi]$, we derive the following expression:
\begin{align}
S[g_{\mu \nu} ,\Phi] \equiv \frac{1}{2} \int \mathrm{d}^5 x \sqrt{-g}
\Bigl[
\partial^{\mu} \Phi \partial_{\mu} \Phi 
\Bigl]\,.
\end{align}
From here, we find the standard Klein-Gordon equation adapted to five dimensions \cite{Crispino:2013pya,Kanti:2014dxa,Pappas:2016ovo,Panotopoulos:2019gtn,Avalos:2023ywb,Gonzalez:2022ote,Rincon:2020cos}
\begin{equation}
\frac{1}{\sqrt{-g}}\partial_{\mu}\left(\sqrt{-g}g^{\mu\nu}\partial_{\nu}\Phi\right) = 0.
\end{equation}
To facilitate the decoupling and subsequent solution of the Klein-Gordon equation, we exploit the symmetries of the metric and introduce the following separation of variables ansatz in spherical coordinates:
\begin{equation} \label{fdc}
\Phi(t, r, \sigma, \theta, \phi) = e^{-i\omega t}\frac{\psi(r)}{r^{3/2}}Y_{\ell m_1 m_2}(\Omega).
\end{equation}
Notice that $\omega$ are the quasinormal frequencies (to be determined), and $Y_{\ell m_1 m_2}(\Omega)$ represents the five-dimensional generalization of the spherical harmonics, which depend on the angular coordinates only \cite{Muller1966-iz}. Thus, to obtain the the quasinormal frequencies $\omega$, we are forced to consider suitable boundary conditions. 
We can split the equation trivially in angular and radial part. In five dimension, the eigenvalues of the differential equation coming from the angular part are $\ell(\ell+2)$ being $\ell$ the angular degree. 
Now, using the method of separation of variables, we obtain a second-order differential equation for the radial coordinate only (taking advantage of the eigenvalues of the angular part). Thus, we have the differential equation given as follow:
\begin{equation}\label{pseudov1}
f(r)^2 \psi''(r) + f(r) f'(r) \psi'(r) + \left[ \omega^2 - f(r) \left( \frac{\ell (\ell + 2)}{r^2} + \frac{3}{2} \frac{f'(r)}{r} + \frac{3}{4} \frac{f(r)}{r^2} \right) \right]\psi(r) =0 \,,
\end{equation}
Even more, we will introduce the so-called "tortoise coordinate" $r_{*}$ defined as usual:
\be
\label{tcd1}
    r_{*}  \equiv  \int \frac{\mathrm{d}r}{f(r)}\,,
\ee
Since the background is asymptotically de~Sitter, the tortoise coordinate does not map the event horizon onto spatial infinity. Instead, $r_{*}$ covers the domain of outer communications, in which the Killing vector $\partial/\partial t$ is timelike: $r_{*}\to-\infty$ corresponds to the event horizon, while $r_{*}\to+\infty$ corresponds to the outer, cosmological-like horizon. All boundary conditions below are imposed on this interval.
to reduce the previous complex differential equation in a more convenient way, namely, in its Schr{\"o}dinger-like form, to obtain
\begin{equation} \label{SLE}
\frac{\mathrm{d}^{2}\psi(r_*)}{\mathrm{d}r_{*}^{2}} + \left[\omega^{2} - V(r_*)\right]\psi(r_*) = 0.
\end{equation}
Should me mention that $V(r)$ is the so-called effective potential defined, in five dimensions, as follow
\begin{equation}\label{poten}
V(r) \equiv f(r)
\Bigg[ 
\frac{\ell (\ell + 2)}{r^2} + \frac{3}{2} \frac{f'(r)}{r} + \frac{3}{4} \frac{f(r)}{r^2}
\Bigg] ,
\end{equation}
where the prime represents the derivative of the radial variable.
Last but not least, the wave equation must be complemented with convenient boundary conditions and, for this case, they are:
\begin{align}
  \Phi \rightarrow \: &e^{+i \omega r_*}, \; \; \; \; \; \;  r_* \rightarrow + \infty
  \quad \text{(outgoing at the cosmological horizon)},
   \\
   \Phi \rightarrow \: &e^{-i \omega r_*}, \; \; \; \; \; \; r_* \rightarrow - \infty
  \quad \text{(ingoing at the event horizon)}.
\label{pbc}
\end{align}
As a concluding remark, given a time dependence of the form $\Phi \sim e^{-i \omega t}$, a frequency with a negative imaginary component physically means a decaying (stable) mode. Conversely, a frequency with a positive imaginary part physically imply a growing (unstable) mode. Thus, the stability/instability is governed by the sign of the imaginary part of quasinormal frequency.

\begin{figure*}
    \centering
    \includegraphics[width=0.325\textwidth]{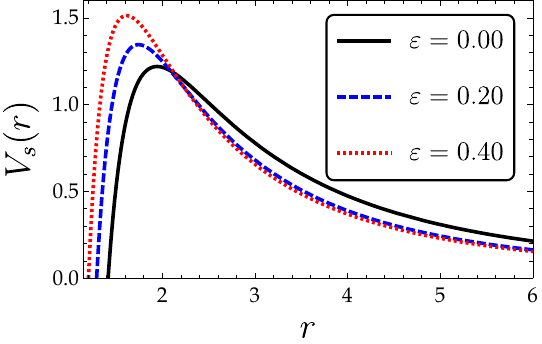}
    \includegraphics[width=0.325\textwidth]{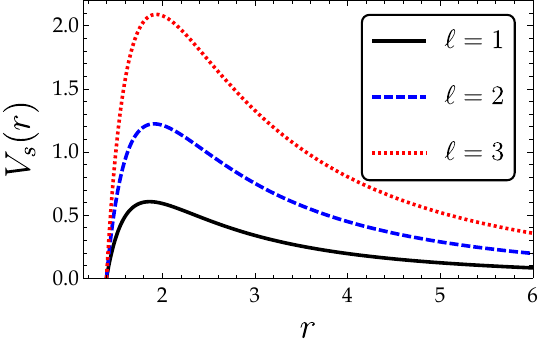}
    \includegraphics[width=0.325\textwidth]{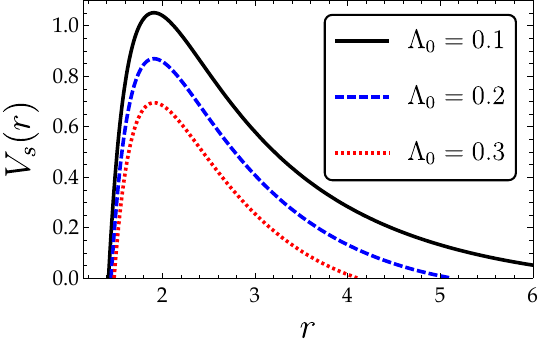}
    \caption{
    Effective potential $V_s(r)$ for scalar perturbations for the scale-dependent Schwarzschild-like black hole solution, varying $\varepsilon$, $\ell $ and $\Lambda_0$, for several fixed values of the remaining parameters.
    {\bf{Left panel:}} Effective potential for 
    $\{ \ell = 1, M_0 =  1, G_0 = 1, \Lambda_0  = +0.01\}$, for three different values of  $\varepsilon$.
    {\bf{Middle panel:}} Effective potential for 
    $\{ M_0 =  1, G_0 = 1, \Lambda_0  = +0.01, \varepsilon = 0.025 \}$, for three different values of  $\ell$.
    {\bf{Right panel:}} Effective potential for 
    $\{ \ell = 2, M_0 =  1, G_0 = 1, \varepsilon = 0.025 \}$, for three different values of  $\varepsilon$.
    }
    \label{Fig:1}
\end{figure*}

In Fig.~\eqref{Fig:1} we show the form of the effective potential barrier $V_s(r)$ for scalar perturbations against the radial coordinate $r$ for different values of the set of parameters $\{ \varepsilon, \ell, \Lambda_0 \}$. 
Notice that when the dimensionless scale-dependent parameter $\varepsilon$ is taken to be zero, we recover the equivalent of the Schwarzschild dS black hole in five dimensions.
Thus, from Fig.~\eqref{Fig:1}, we can read-off the following facts:
\begin{itemize}
    \item {\bf{Left panel}} shows the $V_s(r)$ for fixed $\{ \ell, M_0, G_0, \Lambda_0 \}$ and different values of the dimensionless scale-dependent parameter $\varepsilon$.
    As $\varepsilon$ increases, the peak of the potential increases and shifts to the left, indicating that the effective potential is slightly stronger compared to the general relativity (GR) solution. Additionally, all curves converge as $r$ approaches the outer cosmological-like horizon, with minor differences observed at small radii.
    \item {\bf{Middle panel}} The plot of $V_s(r)$ for fixed $\{ M_0, G_0, \Lambda_0, \varepsilon \}$ and varying angular degree $\ell$ shows that as $\ell$ increases, the potential's maximum rises and shifts slightly to the right.
    \item {\bf{Right panel}} The plot of $V_s(r)$ for fixed $\{ \ell, M_0, G_0, \varepsilon \}$ and varying cosmological constant $\Lambda_0$ reveals that as $\Lambda_0$ increases, the potential's maximum decreases and shifts slightly to the left relative to other curves. At small radial coordinates, the inclusion of $\Lambda_0$ has negligible impact, but as $r$ increases, the effective potential shows slight variations.
\end{itemize}

\subsection{Electromagnetic Perturbations}

Let us summarize the underlying idea, taking advantage of the sum up made for the scalar field case. In five dimensions, 
and by virtue of linear electrodynamics, the electromagnetic field is described by Maxwell’s equations, $\nabla^\mu F_{\mu\nu} = 0$, in which $F_{\mu\nu}$ is the usual field strength tensor. 
Perturbations are decomposed using harmonics on the transverse 3-manifold (typically a 3-sphere for spherical topology). This leads to two distinct types of electromagnetic perturbations:
 i) vector-type modes (derived from vector harmonics on the 3-sphere), and
ii) scalar-type modes (derived from scalar harmonics).
Such a potential depends on i) the dimension of interests, ii) the black hole background and finally, iii) the harmonic mode.
Avoiding unnecessary details, the radial differential equation coming from these perturbations is governed by a Schrödinger-like equation:
\begin{align}
  -\frac{\mathrm{d}^2 \psi}{\mathrm{d}r_*^2} + V(r) \psi = \omega^2 \psi,  
\end{align}
where, as in the scalar case, $\psi(r)$ is the radial wave function, $r_*$ is the tortoise coordinate, $\omega$ is the quasinormal mode frequency, and $V(r)$ is the effective potential (which have a scalar and vector concrete form). 

At this point, let us consider that we are interested in the case $d=5$, therefore, the effective potential for the scalar and vector case takes the following form
\begin{align}
 \label{VSelecto}
    V_S(r) & \equiv f(r)
\Bigg[ 
\frac{\ell (\ell + 2)}{r^2}  + 
\frac{3}{4} \frac{f(r)}{r^2} -
\frac{1}{2} \frac{f'(r)}{r} 
\Bigg],
    \\
     \label{VVelecto}
    V_V(r) & \equiv f(r)
\Bigg[ 
\frac{(\ell + 1)^2}{r^2}  - 
\frac{1}{4} \frac{f(r)}{r^2} +
\frac{1}{2} \frac{f'(r)}{r} 
\Bigg].
\end{align}
The prime denotes differentiation with respect to $ r $. Notably, although not immediately obvious, for $ d = 4 $, the expressions for $ V_S(r) $ and $ V_V(r) $ match the standard electromagnetic potential $ V_{\text{EM}}(r) = f(r) \frac{\ell (\ell + 1)}{r^2} $, where $ \ell $ represents the angular momentum quantum number. This agreement occurs because the last two terms in the effective potential include a $ (d - 4) $ prefactor, which vanishes in four dimensions, canceling those terms. Finally, each potential is associated with a corresponding wave function: $ \psi_S $ for $ V_S(r) $ and $ \psi_V $ for $ V_V(r) $, respectively.

The boundary conditions are also the same that in the scalar case, the reason why we don't need to repeat them (purely ingoing waves at the event horizon $r_H$ and purely outgoing waves at the outer cosmological-like horizon $r_c$). Also, the concrete form of the two effective potentials $\{V_S(r), V_V(r)\}$ are quite similar to the potential for scalar massless perturbations. We then expect a similar behaviour and we will avoid a graphic representation.

\subsection{Dirac Perturbations}

In the next section, we outline the framework for calculating the quasinormal modes of neutral Dirac particles. For a thorough analysis, we assume a spherically symmetric background in $d$-dimensional spacetime, with the specific case of $d=5$ addressed at the end of the derivation (see \cite{Cho:2007zi} for further details).
The metric is then given by the following line element
\begin{align}
    \mathrm{d} s^{2}=-\left|g_{t t}\right| \mathrm{d} t^{2}+g_{r r} \mathrm{d} r^{2} + r^{2}\mathrm{d}\Omega^{2}_{d-2}, 
\end{align}
where $d\Omega^{2}_{d-2}$ represents the metric for the $(d-2)$-dimensional sphere. Also, the $g_{tt}$ and $g_{rr}$ satisfied the Schwarzschild gauge:
\begin{align}
g_{tt} &=  f_0(r) +  \varepsilon \cdot \Bigg(\frac{8 \sqrt{G_0 M_0}}{3 r} - \frac{2r}{3 \sqrt{G_0 M_0}}  \Bigg)  
\equiv 
f(r), 
\\
g_{rr} &= \Bigg[  f_0(r) +  \varepsilon \cdot \Bigg(\frac{8 \sqrt{G_0 M_0}}{3 r} - \frac{2r}{3 \sqrt{G_0 M_0}}  \Bigg)  \Bigg]^{-1}
\equiv 
f(r)^{-1}.
\end{align}
Under the conformal transformation (see \cite{Das:1996we,Gibbons:1993hg} and references therein), we have:
\begin{eqnarray}
g_{\mu\nu} & \rightarrow & \overline{g}_{\mu\nu}=\tilde{\Omega}^{2}g_{\mu\nu} , \\
\psi & \rightarrow & \overline{\psi}=\tilde{\Omega}^{-(d-1)/2}\psi , \\
\gamma^{\mu}\nabla_{\mu}\psi & \rightarrow & \tilde{\Omega}^{(d+1)/2} \overline{\gamma}^{\mu}\overline{\nabla}_{\mu}\overline{\psi} ,
\end{eqnarray}
Now, the metric becomes, after considering $\tilde{\Omega}=1/r$
\begin{align}
\mathrm{d}\overline{s}^{2} = -\frac{1}{r^{2}}f(r)\mathrm{d}t^{2} + \frac{1}{r^{2}} f(r)^{-1}\mathrm{d}r^{2} + \mathrm{d}\Omega^{2}_{d-2} , 
\end{align}
where we have considered $\overline{\psi}=r^{(d-1)/2}\psi$.
Let us now split the $t-r$ part and the $(d-2)$-sphere part in order to decouple the equation and solve for $t-r$. 
Thus, we can write down the Dirac equation for massless particles in the following form:
\begin{align}
\overline{\gamma}^{\mu}\overline{\nabla}_{\mu}\overline{\psi}  =  
\bigg[ 
\left(
\overline{\gamma}^{t}\overline{\nabla}_{t} +
\overline{\gamma}^{r}\overline{\nabla}_{r} \right) \otimes 1
+ 
\overline{\gamma}^{5} \otimes
\left( \overline{\gamma}^{a}
\overline{\nabla}_{a}\right)_{S_{d-2}} 
\bigg] 
\overline{\psi} 
= 0 , 
\end{align}
where as usual $(\overline{\gamma}^{5})^{2}=1$. 
From now on, let us change our notation by omitting the bars.
We shall now let $\chi_{\ell}^{(\pm)}$ be the eigenspinors for the $(d-2)$-sphere (see \cite{Camporesi:1995fb} for further details):
\begin{equation}
\left( \gamma^{a}\nabla_{a} \right)_{S_{d-2}}\chi_{\ell}^{(\pm)} = \pm i \left( \ell + \frac{d-2}{2}\right) \chi_{\ell}^{(\pm)} ,
\end{equation}
Here $\ell = 0, 1, 2, \dots$. Since the eigenspinors are orthogonal, it is always possible to expand $\psi$ as:
\begin{equation}
\psi = \sum_{\ell} \left( \phi_{\ell}^{(+)} \chi_{\ell}^{(+)} + \phi_{\ell}^{(-)} \chi_{\ell}^{(-)} \right) .
\end{equation}
The Dirac equation acquires the simple form:
\begin{equation}\label{eqn:2Ddirac}
\left \{ 
\gamma^{t} \nabla_{t} + \gamma^{r} \nabla_{r} + \gamma^{5} \left[ \pm i \left( \ell + \frac{d-2}{2} \right) \right] \right \} 
\phi_{\ell}^{(\pm)} = 0 ,
\end{equation}
which is the $2$-dimensional Dirac equation.
Let us introduce the explicit choice of the Dirac matrices, according to
\begin{align}
\gamma^{t} &= \frac{r}{\sqrt{f(r)}}(-i\sigma^{3}), 
\\
\gamma^{r} &=  \sqrt{f(r)} 
\ r\sigma^{2} ,
\end{align}
At this point, it is essential to remember that $\sigma^{i}$ are the Pauli matrices, defined as usual, i.e.,:
\begin{equation}
\sigma^{1}=\left(
\begin{array}{cc}
0 & 1 \\ 1 & 0
\end{array}
\right)\ \ \ ,\ \ \ \sigma^{2}=\left(
\begin{array}{cc}
0 & -i \\ i & 0
\end{array}
\right)\ \ \ ,\ \ \ \sigma^{3}=\left(
\begin{array}{cc}
1 & 0 \\ 0 & -1
\end{array}
\right) .
\end{equation}
Also, $\gamma^{5}$ is written in term of the Pauli matrices as follows
\begin{equation}
\gamma^{5} = (-i\sigma^{3})(\sigma^{2}) = - \sigma^{1} .
\end{equation}
The spin connections are then found to be:
\begin{align}
\Gamma_{t} &= \sigma^{1} 
\left( 
\frac{1}{4} r^{2} 
\right) 
\frac{\mathrm{d}}{\mathrm{d}r} \left( \frac{f(r)}{r^{2}} \right) , \\
\Gamma_{r} &= 0 .
\end{align}
The $+$ sign solution is mathematically equivalent to the $-$ sign case, allowing us to focus solely on the positive case. The parameter $\xi$ is then defined in terms of (i) the dimension $d$ and (ii) the angular number $\ell$, as follows:
\begin{align}
    \xi \equiv \ell + \frac{1}{2}(d-2), 
\end{align}
with $\xi$ running $ +3/2, +5/2,...$ with $d=5$. 
Also note that, although the value of the variable $\xi$ could be either positive or negative, we will focus on the positive part of the set for simplicity. 
Taking the last facts into account, the Dirac equation is reduced to:
\begin{align}
& \left\{ \frac{r}{\sqrt{f(r)}}(-i\sigma^{3}) \left[
\frac{\partial}{\partial t} + 
\sigma^{1} 
\left( 
\frac{1}{4} r^{2} 
\right) 
\frac{\mathrm{d}}{\mathrm{d}r} \left( \frac{f(r)}{r^{2}} \right)
\right] + 
\sqrt{f(r)} 
r \sigma^{2}
\frac{\partial}{\partial r} + (-\sigma^{1})(i) \left( \xi \right) \right\} \phi_{\ell}^{(+)} = 0  
\end{align}
Putting spatial and temporal part separately, we finally have a first order partial differential equation for $\phi_{\ell}^{(+)}$, i.e., 
\begin{align}
\sigma^{2} 
\left( 
\sqrt{f(r)} \
r
\right) 
\left[ \frac{\partial}{\partial r} + \frac{r}{2\sqrt{f(r)}}
\frac{\mathrm{d}}{\mathrm{d}r} \left( \frac{\sqrt{f(r)}}{r} \right) \right]
\phi_{\ell}^{(+)} - i \sigma^{1} \xi 
\phi_{\ell}^{(+)} = i \sigma^{3} \left( \frac{r}{\sqrt{f(r)}} \right)
\frac{\partial \phi_{\ell}^{(+)}}{\partial t} . &
\end{align}
Considering solutions of the form:
\begin{equation}
\phi_{\ell}^{(+)} = \left( \frac{\sqrt{f(r)}}{r} \right)^{-1/2} e^{-i \omega t} \left(
\begin{array}{c}
i \tilde{G}(r) \\ 
 F(r)
\end{array}
\right) ,
\end{equation}
The Dirac equation is:
\begin{equation}
\sigma^{2} \left( \sqrt{f(r)} \ r \right) \left(
\begin{array}{c}
i\frac{\mathrm{d}\tilde{G}(r)}{\mathrm{d}r} \\ 
\ \frac{\mathrm{d}F(r)}{\mathrm{d}r}
\end{array}
\right) -i \sigma^{1} \xi \left(
\begin{array}{c}
i \tilde{G}(r) 
\\ 
F(r)
\end{array}
\right) = \sigma^{3} \omega \left( \frac{r}{\sqrt{f(r)}} \right) \left(
\begin{array}{c}
i \tilde{G}(r) 
\\ 
F(r)
\end{array}
\right) .
\end{equation}
Finally, by treating the components separately, we derive a set of first-order differential equations for the functions $\{F(r), \tilde{G}(r)\}$ as follows:
\begin{eqnarray}
f(r) \frac{\mathrm{d}\tilde{G}(r)}{\mathrm{d}r} 
- 
\Bigg [ \frac{\sqrt{f(r)}}{r} \xi \Bigg ] \tilde{G}(r) & = & + \omega F(r) ,  \label{Eq1}
\\ 
f(r) \frac{\mathrm{d}F(r)}{\mathrm{d}r} 
+ 
\Bigg [ \frac{\sqrt{f(r)}}{r} \xi \Bigg ] F(r) & = & - \omega \tilde{G}(r) . \label{Eq2}
\end{eqnarray}
For simplicity, let us introduce: 
 i) the tortoise coordinate $r_*$, and 
ii) the auxiliary function $W(r)$, 
as follows
\begin{align}
r_*(r) &\equiv \int ^{r}\mathrm{d}\bar{r} 
\Bigl[ 
f(\bar{r}) 
\Bigl]^{-1} ,
\\
W(r) &= \frac{\xi \sqrt{f(r)}}{r}
\end{align}
and the coupled differential equations \eqref{Eq1} and \eqref{Eq2} are rewritten as
\begin{align}
    \Bigg[
    \frac{\mathrm{d}}{\mathrm{d}r_{*}} - W(r)
    \Bigg] \tilde{G}(r) &= + \omega F(r)
    \\
    \Bigg[
    \frac{\mathrm{d}}{\mathrm{d}r_{*}} + W(r)
    \Bigg] F(r) &= - \omega \tilde{G}(r)
\end{align}
The set of first-order differential equations for $\tilde{G}(r)$ and $F(r)$ can be trivially decoupled to obtain two Schrodinger-like differential equations, i.e., 
\begin{align}
\frac{\mathrm{d}^2F}{\mathrm{d}{r_{*}}^2} + [\omega^2 - V_{-}] F & =  0 , \label{SL1}
\\
\frac{\mathrm{d}^2\tilde{G}}{\mathrm{d}{r_{*}}^2} + [\omega^2 - V_{+}] \tilde{G} & =  0 , \label{SL2}
\end{align}
with the potentials defined by 
\begin{equation}
V_{\pm} = W^2 \pm \frac{\mathrm{d}W}{\mathrm{d}r_{*}} .
\end{equation}
In General Relativity, the effective potentials $ V_{+} $ and $ V_{-} $ produce identical spectra (isospectrality) for Schwarzschild and Reissner-Nordström black holes. However, this property does not hold in several cases, such as: 
  i) alternative theories of gravity, 
 ii) higher-dimensional gravity, 
iii) near-extremal black holes, and others. 
Given that we are studying a black hole solution beyond General Relativity and in higher dimensions, isospectrality is expected to be broken. Typically, the resulting spectral differences are small, so we will focus on a single case. In the following, we adopt the positive sign for the effective potential and express it in the standard radial coordinate as:
\begin{align}
V_{+} &= f(r) 
\Bigg[
\frac{\xi ^2}{r^2} + 
\frac{\xi  f'(r)}{2 r \sqrt{f(r)}}-\frac{\xi  \sqrt{f(r)}}{r^2}
\Bigg].
\end{align}
Note that a convenient advantage of performing the conformal transformation from the initial stage is that the effective potential remains unchanged regardless of the dimension under consideration, as the dimension solely appears in the effective potential via the parameter  $\xi$. Thus, the effective potential maintains the same functional form for arbitrary higher dimensions.

Consistent with the scalar and electromagnetic cases, the boundary conditions are imposed on the domain of outer communications: purely ingoing waves at the event horizon and purely outgoing waves at the outer, cosmological-like horizon (no incoming radiation from the cosmological horizon and no outgoing radiation from the event horizon). Accordingly, the suitable boundary conditions for the Schrödinger-like equations \eqref{SL1} and \eqref{SL2} are:
\begin{align}
   \{F, \tilde{G}\} \rightarrow \: &e^{+i \omega r_*}, \; \; \; \; \; \;  r_* \rightarrow + \infty
   \quad \text{(outgoing at the cosmological horizon)},
   \\
   \{F, \tilde{G}\} \rightarrow \: &e^{-i \omega r_*}, \; \; \; \; \; \; r_* \rightarrow - \infty
   \quad \text{(ingoing at the event horizon)}.
\end{align}
As previously discussed, for $\psi \sim e^{-i \omega t}$, a frequency with a negative imaginary part corresponds to a stable mode, whereas a frequency with a positive imaginary part indicates an  unstable mode.

\begin{figure*}
    \centering
    \includegraphics[width=0.325\textwidth]{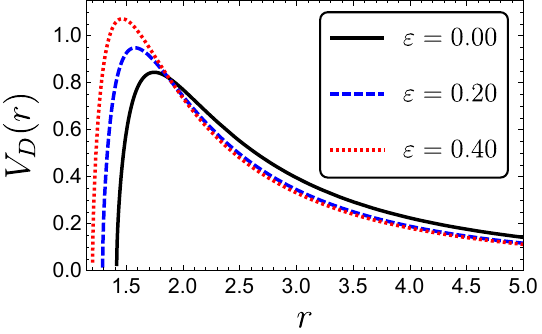}
    \includegraphics[width=0.325\textwidth]{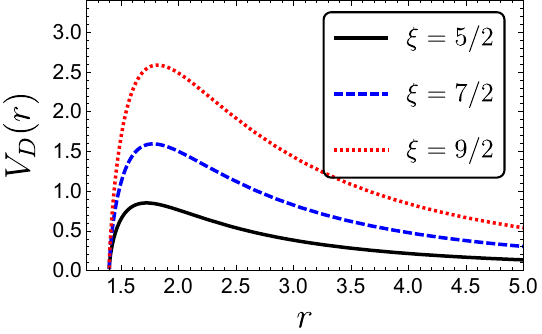}
    \includegraphics[width=0.325\textwidth]{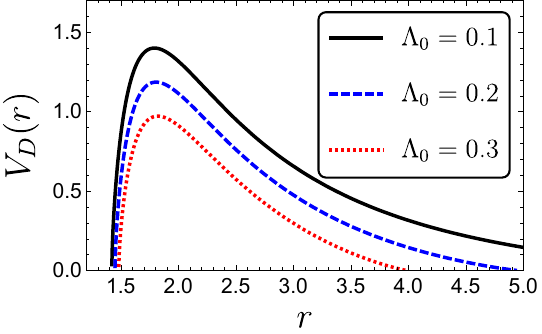}
    \caption{
    Effective potential for massless Dirac perturbations for the 5-dimensional scale-dependent black hole. We show $V_D(r)$ varying $\varepsilon$, $\xi$ and $\Lambda_0$ for several different parameter values involved.
    {\bf{Left panel:}} Effective potential fixing $\xi=5/2$, $\Lambda_0=0.01$ and $M_0 = 1 = G_0$, varying $\varepsilon$.
    {\bf{Middle panel:}} Effective potential fixing $\varepsilon=0.025$, $\Lambda_0=0.01$ and $M_0 = 1 = G_0$ varying $\xi$.
    {\bf{Right panel:}} Effective potential fixing $\xi=7/2$, $\varepsilon=0.025$ and $M_0 = 1=G_0$ varying $\Lambda_0$.
    }
    \label{Fig:2}
\end{figure*}

Once again, we plot the effective potential, $V_D(r)$, for the five-dimensional case against the radial coordinate, for various values of the parameters, $\{\varepsilon, \xi, \Lambda_0\}$, with $M_0 = 1$ and $G_0 = 1$ fixed. For comparison purposes, we also include the classical case with $\varepsilon = 0$.
We have considered only three cases:  
\begin{itemize}
    \item Fig.~\eqref{Fig:2} (left panel) shows, for three different values of $\varepsilon$, how the effective potential changes. Thus, when the scale-dependent parameter increases, the effective potential decreases, meaning that the inclusion of quantum corrections affects significantly $V_D(r)$.
    \item Fig.~\eqref{Fig:2} (middle panel) shows,  for three different values of $\xi$,
    the evolution of the effective potential, keeping the remaining parameters fixed.
   The figure reveals that as $\xi$ increases, the maximum of the potential increases, at the same time as it shifts to the right. In addition, the effective potential tends to overlap quickly near the event horizon and near the outer cosmological-like horizon.
    \item Fig.~\eqref{Fig:2} (right panel) shows,  for three different values of $\Lambda_0$, the effective potential. As expected a small modification to the cosmological constant does not significantly alter the concrete form of $V_D(r)$.
\end{itemize}

\section{Method: WKB approximation}
\label{sec:meth}
In the context of the black hole background introduced above, we employ the well-known WKB approach \cite{Konoplya:2011qq,Konoplya:2003ii,Konoplya:2019hlu,Konoplya:2020bxa,Konoplya:2024lir,Konoplya:2023moy,Koch:2025gaw,Pantig:2025eda,Rincon:2025buq,Balart:2024rtj,Hamil:2024ppj,Lutfuoglu:2025hjy,Lutfuoglu:2025hwh,Lambiase:2024lvo,Gogoi:2023fow,Baruah:2023rhd,Lutfuoglu:2025ljm,Konoplya:2023aph,Konoplya:2022hll,Churilova:2019sah,Turimov:2019afv}, as the effective potential exhibits a suitable barrier form. This method has been investigated in detail on several occasions, which is why we mention only the minimum context required to keep the paper self-contained.

\begin{figure*}
    \centering
    \includegraphics[width=0.75\textwidth]{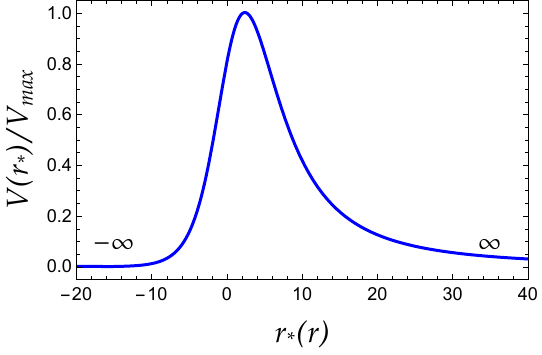}
    \caption{
Illustrative representation of the effective potential convenient to the implementation of the WKB method.
    }
    \label{Fig:3}
\end{figure*}

The WKB approximation is a powerful semi-analytical method used to calculate the quasinormal mode (QNM) spectrum in scenarios where the effective potential has a clearly defined peak in the domain of outer communications, between the event horizon and the outer cosmological-like horizon (see Fig.~\eqref{Fig:3} for illustration). 
Given the natural correspondence between the perturbation equations in black hole spacetimes and the one-dimensional Schrödinger-like equation with a potential barrier, studying QNM frequencies is equivalent to solving a quantum mechanical scattering problem, thus reducing the complexity of the issue to a more straightforward and familiar situation. The key issue is the boundary conditions on the domain of outer communications: (i) purely ingoing waves must be imposed at the event horizon ($r_{*}\to-\infty$), and (ii) purely outgoing waves must be imposed at the outer, cosmological-like horizon ($r_{*}\to+\infty$).

The underlying idea is as follows: regardless of the type of perturbation, the aim is to split the perturbed equations into radial and angular parts using the method of separation of variables. The angular part then enters the radial Schrödinger-like equation via its angular eigenvalue, reducing the situation to the resolution of a second-order differential equation subject to an effective barrier, which is often non-trivial, and supplemented by suitable boundary conditions, as mentioned a few lines ago.
Thus, the equation might be written as follow
\begin{equation}\label{wkb-schro}
   \Bigg[
    \frac{\mathrm{d}^{2}}{\mathrm{d}x^{2}} + \Bigl(\omega^{2} - V(x) \Bigl)
   \Bigg] \Psi(x)= 0,
\end{equation}
where the coordinate $x$ is linked to the tortoise coordinate $r_{*}$ (or radial coordinate), which maps the event horizon and the outer cosmological-like horizon to $-\infty$ and $+\infty$, respectively. In addition, $V(x)$ is the effective potential (which varies depending of the applied perturbation and also the dimensionality of the spacetime), and finally $\omega$ is the so-called quasinormal frequency.

Notice that there are three well-defined regions of interest here (in what follows we will use the radial coordinate as variable). Since the quasinormal frequency $\omega$ is complex, the inequalities that distinguish oscillatory from evanescent regions are written in terms of the real quantity $|\omega|^{2}-V(r)$, rather than $\omega^{2}-V(r)$.
\begin{itemize}
    \item {\bf{Region I}}: Near the event horizon ($r \to r_H$, $r_* \to -\infty$), where $r<r_1$, $|\omega|^{2} - V(r) > 0$, and the solution is oscillatory (ingoing waves).
     \item {\bf{Region II}}: Between the turning points $r_1 < r < r_2$ (where $|\omega|^{2} = V(r)$), with $|\omega|^{2} - V(r) < 0$, and the solution is exponential (classically forbidden).
      \item {\bf{Region III}}: Near the outer, cosmological-like horizon ($r \to r_c$, $r_* \to +\infty$), where $r > r_2$, $|\omega|^{2} - V(r) > 0$, and the solution is oscillatory (outgoing waves).
\end{itemize}
To the correct application of the method, it is required matching the wave functions across the turning points $r_1$ and $r_2$ to ensure continuity and derive the QNM quantization condition.

The effective potential vanishes at spatial infinity in the asymptotically flat case. In the asymptotically de~Sitter geometry considered here, $V(r)$ vanishes both at the event horizon and at the outer cosmological-like horizon, and it forms a barrier in the domain of outer communications.
Quasinormal mode frequencies $\omega_{n\ell}$ are obtained by solving equation~(\ref{wkb-schro}) with purely outgoing waves at that cosmological horizon ($r_{*} \to +\infty$) and purely ingoing waves at the event horizon ($r_{*} \to -\infty$), then matching the solutions in the intermediate region. Here, $n$ is the overtone number and $\ell$ the multipole number. To ensure numerical convergence and stability, one should restrict to $\ell \geq n$.
The quantization condition requires the appearance of a pole in the Gamma function, specifically $\Gamma(-n) = \infty$ (where $\Gamma$ is the ordinary Gamma function and $n = 0, 1, 2, \dots$). This condition ultimately yields the following expression for the quasinormal mode frequencies:
\begin{equation}\label{wkb-formula}
   n + \frac{1}{2} = i \frac{\omega^2 - V_0}{\sqrt{-2V_0''}} - \sum_{j=2}^{6} \Pi_j.
\end{equation}
The symbols appearing in the expression are defined as follows:
i) $V_0$ is the height (maximum value) of the effective potential,
ii) $\Pi_j \equiv \Pi_j(n, V_0', V_0'', \dots)$ represent coefficients that depend on the overtone number $n$ and on the value of the effective potential together with its derivatives (denoted by primes) evaluated at the location of the maximum and finally, 
iii) $\omega$ is the quasinormal mode frequency.
The quasinormal frequencies are obtained once the overtone number and the effective potential are provided, by solving the formula Eq.~(\ref{wkb-formula}) for the corresponding $\omega$.

The WKB approximation was originally implemented in standard quantum mechanics to compute the energy levels of quantum systems. The same idea applies in black-hole physics; however, the demand for higher accuracy requires us to go beyond the eikonal approximation resulting in the inclusion of higher-order corrections \cite{Konoplya:2011qq}. Consequently, the method has been extended to higher orders: up to third order in early studies \cite{Schutz:1985km,Will:1985gp,Iyer:1986np}, sixth order in \cite{Konoplya:2003ii}, and as high as thirteenth order in \cite{Matyjasek:2017psv}.
It is also worth noting that, even under the assumption $\ell \gg n$ (or $\ell \geq n$ with sufficiently large $\ell$), it is not trivial to predict a priori the order at which the WKB approximation yields the best accuracy. In practice, the third- and sixth-order versions typically provide highly accurate results for quasinormal modes and other observables. For this reason, the present study focuses primarily on these two orders.
When the order of  WKB expansion increase, the  functions $\Pi_j$ becomes larger and more complex (see the expressions in \cite{Iyer:1986vv,Konoplya:2003ii,Matyjasek:2017psv}).

\section{Numerical solutions and Discussion}
\label{sec:nume}

After a short revision of the WKB method, we are in position to discuss the results obtained in each background. 

\subsection{Scalar Perturbations}

\begin{figure*}
    \centering
    \includegraphics[width=0.49\textwidth]{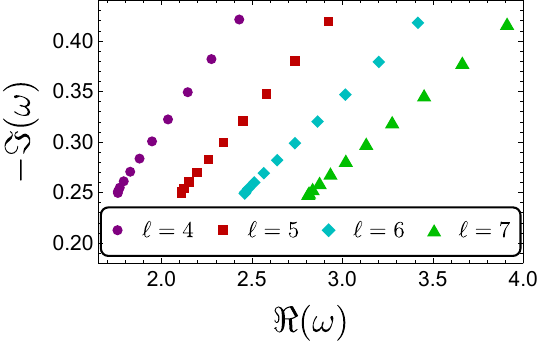}
    \
    \includegraphics[width=0.49\textwidth]{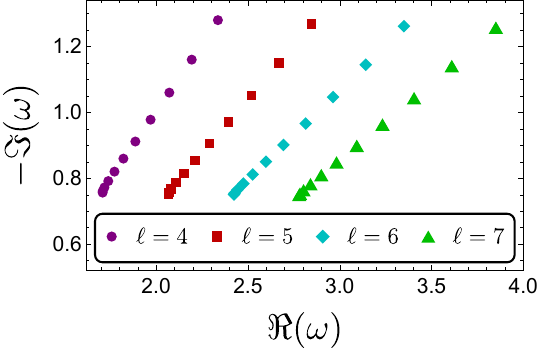}
    \caption{
     Quasinormal modes of scalar perturbations for different values of the scale-dependent parameter $\varepsilon$ ($0.0 \leq \varepsilon \leq 0.5$) and the angular number $\ell$, keeping fixed $\{ M_0 =  1, G_0 = 1, \Lambda_0  = +0.01\}$.
    {\bf{Left panel:}} Quasinormal modes for the fundamental mode $(n=0)$. 
    {\bf{Right panel:}} Quasinormal modes for the first exited mode $(n=1)$. 
    }
    \label{Fig:4}
\end{figure*}

\begin{figure} 
\begin{center}
\includegraphics[width=0.49\linewidth]{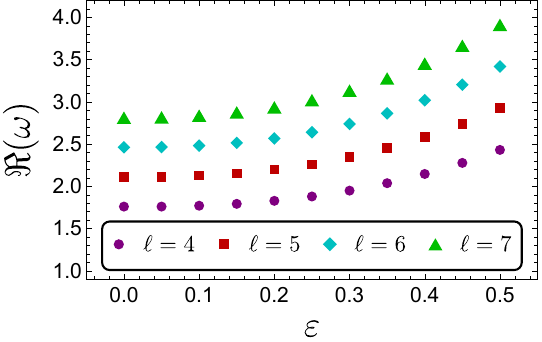} 
\includegraphics[width=0.49\linewidth]{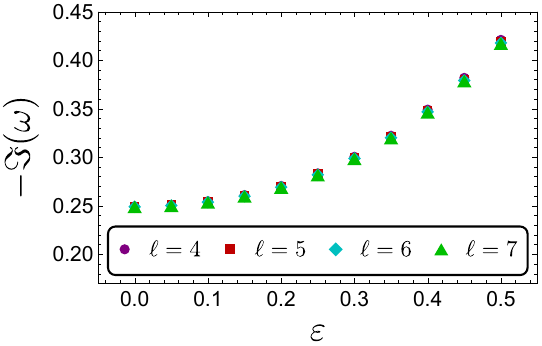}
\\
\vspace{0.5cm}
\includegraphics[width=0.49\linewidth]{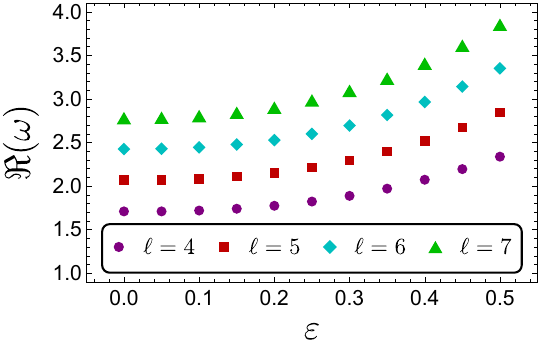} 
\includegraphics[width=0.49\linewidth]{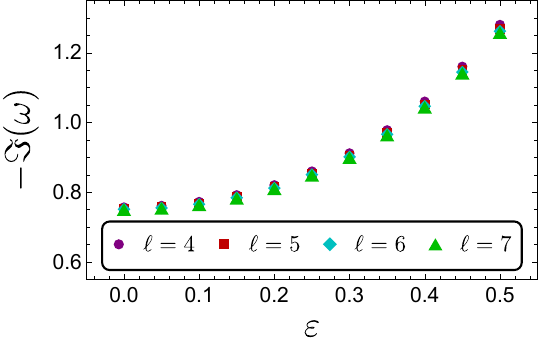}
\caption{
Real/imaginary part of Scalar quasinormal frequencies, $\omega_R$/$\omega_{I}$, against the scale-dependent parameter $\varepsilon$. 
{\bf{Top Left Panel:}} $\text{Re}(\omega)$ vs $\varepsilon$ for $M_0=1$, $G_0=1$ and $\Lambda_0=0.01$, varying $\ell$ from 4 to 7  for the fundamental mode $(n=0)$.
{\bf{Top Right Panel:}} $-\text{Im}(\omega)$ vs $\varepsilon$ for $M_0=1$, $G_0=1$ and $\Lambda_0=0.01$, varying $\ell$ from 4 to 7  for the fundamental mode $(n=0)$.
{\bf{Down Left Panel:}} $\text{Re}(\omega)$ vs $\varepsilon$ for $M_0=1$, $G_0=1$ and $\Lambda_0=0.01$, varying $\ell$ from 4 to 7 for the first exited mode $(n=1)$.
{\bf{Down Right Panel:}} $-\text{Im}(\omega)$ vs $\varepsilon$ for $M_0=1$, $G_0=1$ and $\Lambda_0=0.01$, varying $\ell$ from 4 to 7 for the first exited mode $(n=1)$.
}
\label{Fig:5}
\end{center}
\end{figure}

Figure \eqref{Fig:4}, consisting of two panels, illustrates the behavior of the fundamental quasinormal modes (left panel) and the first excited modes (right panel) for varying values of the extra parameter $\varepsilon$. Note that this dimensionless parameter, derived from the scale-dependent formalism, is expected to be significantly small. 
Figure \eqref{Fig:5} shows the direct impact of a change in the scale-dependent parameter $\varepsilon$ for the real and imaginary parts of the fundamental mode ($n=0$) in the first row and the first excited mode ($n=1$) in the second row.

\begin{table}[ph!]
\centering
\caption{
Quasinormal frequencies for massless scalar perturbations (varying $\ell$, $n$ and $\varepsilon$) fixing $M_0=1, G_0=1, \Lambda_0=0.01$ for the model considered in this work using the WKB approximation.
}
{
\resizebox{0.85\columnwidth}{!}
    {
\begin{tabular}{c|c|cccc} 
\toprule
$\varepsilon$  &$n$  &  $\ell=4$ & $\ell=5$ & $\ell=6$ & $\ell=7$ 
\\ 
\colrule
\hline
\rowcolor{purple!30} 0.00 & 0 & 1.759631-0.249496 i & 2.110190-0.249140 i & 2.460934-0.248924 i & 2.811791-0.248785 i \\
\rowcolor{purple!30} 0.00 & 1 & 1.708752-0.756227 i & 2.067505-0.752787 i & 2.424202-0.750714 i & 2.779569-0.749370 i \\
\rowcolor{purple!30} 0.00 & 2 & 1.610650-1.286714 i & 1.984147-1.272867 i & 2.351956-1.264530 i & 2.715917-1.259132 i \\
\rowcolor{purple!30} 0.00 & 3 & 1.473672-1.856982 i & 1.864797-1.820734 i & 2.247027-1.798741 i & 2.622658-1.784458 i \\
\botrule
\hline
0.05 & 0 & 1.759792-0.250980 i & 2.111865-0.250483 i & 2.463934-0.250183 i & 2.815998-0.249989 i \\
0.05 & 1 & 1.708891-0.760723 i & 2.069113-0.756845 i & 2.427121-0.754510 i & 2.783692-0.752996 i \\
0.05 & 2 & 1.610827-1.294324 i & 1.985656-1.279717 i & 2.354730-1.270919 i & 2.719883-1.265222 i \\
0.05 & 3 & 1.474062-1.867779 i & 1.866233-1.830472 i & 2.249621-1.807805 i & 2.626406-1.793077 i \\
\botrule
\hline
0.10 & 0 & 1.769975-0.254700 i & 2.125416-0.254075 i & 2.480685-0.253699 i & 2.835841-0.253455 i \\
0.10 & 1 & 1.718539-0.772014 i & 2.082182-0.767705 i & 2.443441-0.765114 i & 2.803149-0.763436 i \\
0.10 & 2 & 1.619519-1.313571 i & 1.997804-1.298102 i & 2.370208-1.288792 i & 2.738577-1.282768 i \\
0.10 & 3 & 1.481588-1.895561 i & 1.877125-1.856809 i & 2.263896-1.833259 i & 2.643986-1.817962 i \\
\botrule
\hline
0.15 & 0 & 1.791993-0.261051 i & 2.153035-0.260304 i & 2.513750-0.259855 i & 2.874257-0.259564 i \\
0.15 & 1 & 1.739398-0.791305 i & 2.108804-0.786546 i & 2.475639-0.783692 i & 2.840801-0.781845 i \\
0.15 & 2 & 1.638215-1.346537 i & 2.022496-1.330043 i & 2.400700-1.320132 i & 2.774711-1.313728 i \\
0.15 & 3 & 1.497455-1.943405 i & 1.899112-1.902692 i & 2.291921-1.877966 i & 2.677892-1.861922 i \\
\botrule
\hline
0.20 & 0 & 1.827767-0.270473 i & 2.197040-0.269605 i & 2.565848-0.269083 i & 2.934361-0.268746 i \\
0.20 & 1 & 1.773249-0.819950 i & 2.151179-0.814697 i & 2.526329-0.811552 i & 2.899667-0.809521 i \\
0.20 & 2 & 1.668439-1.395583 i & 2.061700-1.377825 i & 2.448613-1.367177 i & 2.831122-1.360308 i \\
0.20 & 3 & 1.522853-2.014830 i & 1.933847-1.971473 i & 2.335814-1.945173 i & 2.730698-1.928131 i \\
\botrule
\hline
0.25 & 0 & 1.879242-0.283455 i & 2.259781-0.282457 i & 2.639730-0.281860 i & 3.019305-0.281473 i \\
0.25 & 1 & 1.821870-0.859450 i & 2.211510-0.853622 i & 2.598133-0.850142 i & 2.982788-0.847898 i \\
0.25 & 2 & 1.711659-1.463344 i & 2.117344-1.443985 i & 2.516327-1.432404 i & 2.910631-1.424948 i \\
0.25 & 3 & 1.558864-2.113797 i & 1.982890-2.066900 i & 2.397613-2.038495 i & 2.804910-2.020120 i \\
\botrule
\hline
0.30 & 0 & 1.948290-0.300522 i & 2.343517-0.299380 i & 2.738042-0.298696 i & 3.132122-0.298255 i \\
0.30 & 1 & 1.886933-0.911433 i & 2.291886-0.904911 i & 2.693549-0.901026 i & 3.093064-0.898524 i \\
0.30 & 2 & 1.769196-1.552701 i & 2.191197-1.531289 i & 2.606051-1.518510 i & 3.015880-1.510296 i \\
0.30 & 3 & 1.606369-2.244681 i & 2.047582-2.193092 i & 2.479130-2.161889 i & 2.902807-2.141735 i \\
\botrule
\hline
0.35 & 0 & 2.036614-0.322231 i & 2.450300-0.320920 i & 2.863184-0.320138 i & 3.275558-0.319632 i \\
0.35 & 1 & 1.969914-0.977631 i & 2.394162-0.970252 i & 2.814805-0.965865 i & 3.233089-0.963044 i \\
0.35 & 2 & 1.842115-1.666754 i & 2.284745-1.642696 i & 2.719683-1.628366 i & 3.149168-1.619172 i \\
0.35 & 3 & 1.665957-2.412248 i & 2.128930-2.354496 i & 2.581814-2.319603 i & 3.026274-2.297097 i \\
\botrule
\hline
0.40 & 0 & 2.145667-0.349161 i & 2.581877-0.347649 i & 3.017194-0.346747 i & 3.451946-0.346165 i \\
0.40 & 1 & 2.072005-1.059861 i & 2.519859-1.051406 i & 2.963744-1.046390 i & 3.405023-1.043168 i \\
0.40 & 2 & 1.931161-1.808783 i & 2.399097-1.781320 i & 2.858696-1.764989 i & 3.312321-1.754525 i \\
0.40 & 3 & 1.737861-2.621600 i & 2.227515-2.555848 i & 2.706633-2.516136 i & 3.176667-2.490545 i \\
\botrule
\hline
0.45 & 0 & 2.276609-0.381907 i & 2.739633-0.380150 i & 3.201688-0.379103 i & 3.663128-0.378428 i \\
0.45 & 1 & 2.194076-1.159995 i & 2.670114-1.150187 i & 3.141760-1.144378 i & 3.610514-1.140651 i \\
0.45 & 2 & 2.036706-1.982224 i & 2.534927-1.950401 i & 3.024069-1.931497 i & 3.506613-1.919398 i \\
0.45 & 3 & 1.821919-2.878181 i & 2.343444-2.802135 i & 2.854017-2.756190 i & 3.354743-2.726600 i \\
\botrule
\hline
0.50 & 0 & 2.430300-0.421069 i & 2.924587-0.419013 i & 3.417842-0.417789 i & 3.910442-0.416999 i \\
0.50 & 1 & 2.336666-1.279949 i & 2.845662-1.268442 i & 3.349786-1.261636 i & 3.850683-1.257273 i \\
0.50 & 2 & 2.158764-2.190642 i & 2.692471-2.153277 i & 3.216275-2.131091 i & 3.732749-2.116902 i \\
0.50 & 3 & 1.917621-3.187697 i & 2.476354-3.098572 i & 3.023839-3.044655 i & 3.560643-3.009923 i \\
\botrule
\hline
\end{tabular} 
\label{table:First set}
    }
}
\end{table}

Based on our results, we conclude that this 5-dimensional black hole solution is stable against massless scalar perturbations. This stability arises because the imaginary part of the frequencies is negative in all cases studied (see Table \eqref{table:First set} to see frequencies). Furthermore, the figures show that both the real and (minus) imaginary parts of the frequencies increase as the parameter $\varepsilon$ increases, indicating a more pronounced correction effect.
Specifically, the real part represents the angular frequency of the oscillation. As $\varepsilon$ increases, the black hole's oscillation frequency decreases compared to the classical spectrum.
Similarly, the imaginary part encodes the damping due to energy loss through gravitational radiation escaping across the outer cosmological-like horizon. As the scale-dependent parameter $\varepsilon$ increases, the damping rate, $-\text{Im}(\omega) > 0$, which is directly related to the oscillation amplitude, also decreases relative to the classical solution.
Thus, the inclusion of quantum corrections introduces significant deviations from the classical spectrum, even for small values of the scale-dependent parameter $\varepsilon$.

\subsection{Electromagnetic Perturbations}

\begin{figure*}
    \centering
    \includegraphics[width=0.49\textwidth]{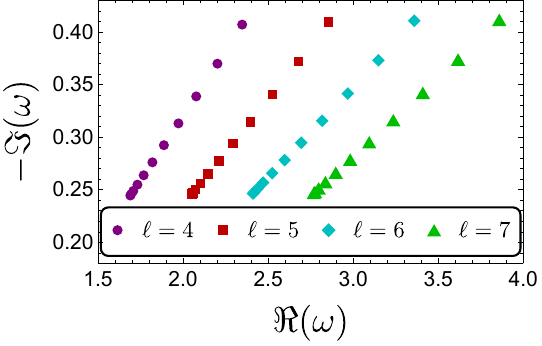} \
    \includegraphics[width=0.49\textwidth]{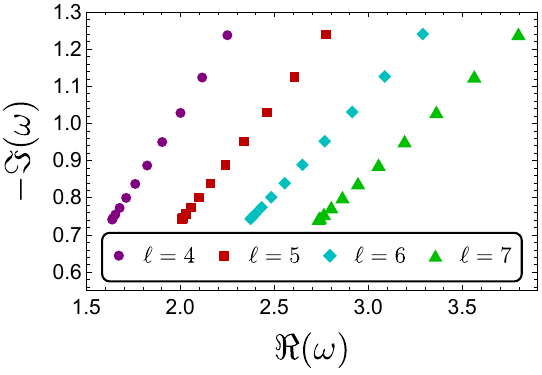} 
    \\
    \vspace{0.5cm}
    \includegraphics[width=0.49\textwidth]{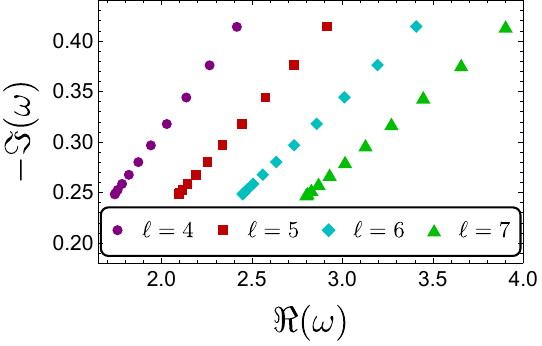} \
    \includegraphics[width=0.49\textwidth]{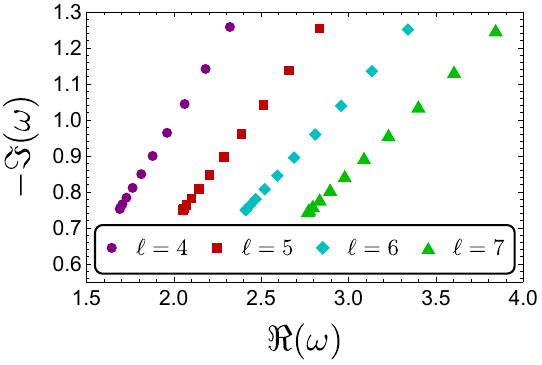}
    \caption{
    Quasinormal modes of Electromagnetic perturbations for different values of the scale-dependent parameter $\varepsilon$ ($0.0 \leq \varepsilon \leq 0.5$) and the angular number $\ell$, keeping fixed $\{ M_0 =  1, G_0 = 1, \Lambda_0  = +0.01\}$. We have considered the two potentials: the  $V_S(r)$ (for the scalar case) and $V_V(r)$ (for the vector case).
    {\bf{Top-left panel:}} Quasinormal modes for the fundamental mode $(n=0)$ for $V_S(r)$.
    {\bf{Top-right panel:}} Quasinormal modes for the first exited mode $(n=1)$ for $V_S(r)$.
    {\bf{Bottom-left panel:}} Quasinormal modes for the fundamental mode $(n=0)$ for $V_V(r)$.
    {\bf{Bottom-right panel:}} Quasinormal modes for the first exited mode $(n=1)$ for $V_V(r)$.
    }
    \label{Fig:6}
\end{figure*}

\begin{figure} 
\begin{center}
\includegraphics[width=0.49\linewidth]{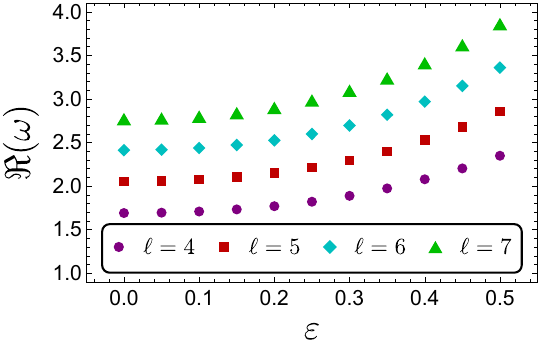} 
\includegraphics[width=0.49\linewidth]{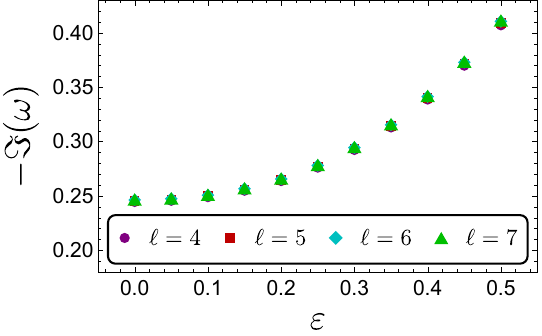}
\\
\vspace{0.5cm}
\includegraphics[width=0.49\linewidth]{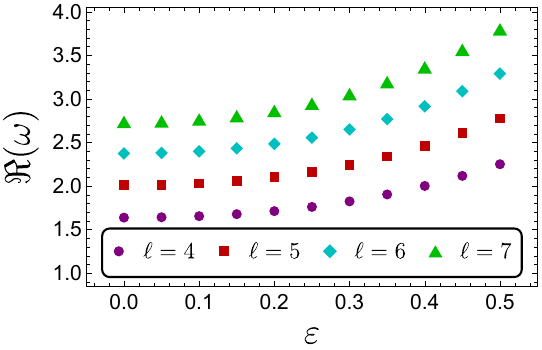} 
\includegraphics[width=0.49\linewidth]{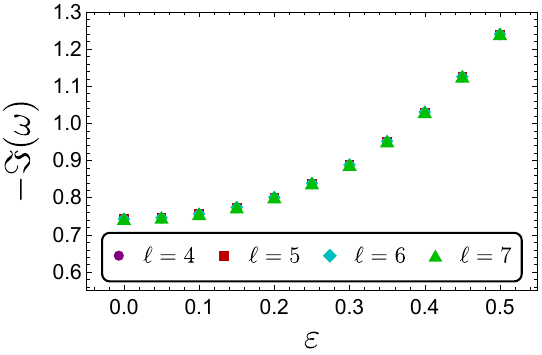}
\caption{
Real/imaginary part of Electromagnetic quasinormal frequencies, $\omega_R$/$\omega_{I}$, against the scale-dependent parameter $\varepsilon$ for the scalar effective potential $V_S(r)$.
{\bf{Top Left Panel:}} $\text{Re}(\omega)$ vs $\varepsilon$ for $M_0=1$, $G_0=1$ and $\Lambda_0=0.01$, varying $\ell$ from 4 to 7  for the fundamental mode $(n=0)$.
{\bf{Top Right Panel:}} $-\text{Im}(\omega)$ vs $\varepsilon$ for $M_0=1$, $G_0=1$ and $\Lambda_0=0.01$, varying $\ell$ from 4 to 7  for the fundamental mode $(n=0)$.
{\bf{Down Left Panel:}} $\text{Re}(\omega)$ vs $\varepsilon$ for $M_0=1$, $G_0=1$ and $\Lambda_0=0.01$, varying $\ell$ from 4 to 7 for the first exited mode $(n=1)$.
{\bf{Down Right Panel:}} $-\text{Im}(\omega)$ vs $\varepsilon$ for $M_0=1$, $G_0=1$ and $\Lambda_0=0.01$, varying $\ell$ from 4 to 7 for the first exited mode $(n=1)$.
}
\label{Fig:7}
\end{center}
\end{figure}

\begin{figure} 
\begin{center}
\includegraphics[width=0.49\linewidth]{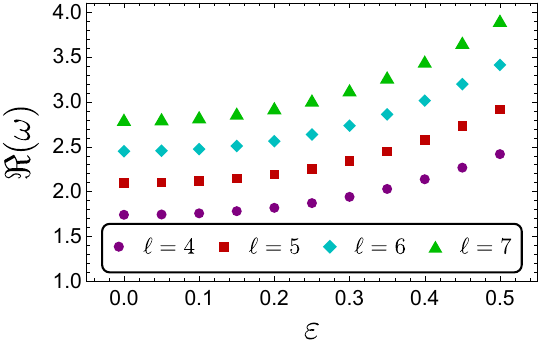} 
\includegraphics[width=0.49\linewidth]{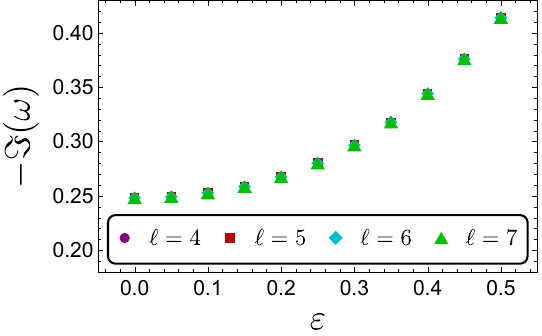}
\\
\vspace{0.5cm}
\includegraphics[width=0.49\linewidth]{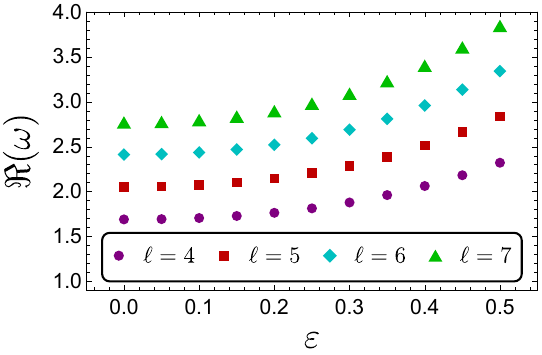} 
\includegraphics[width=0.49\linewidth]{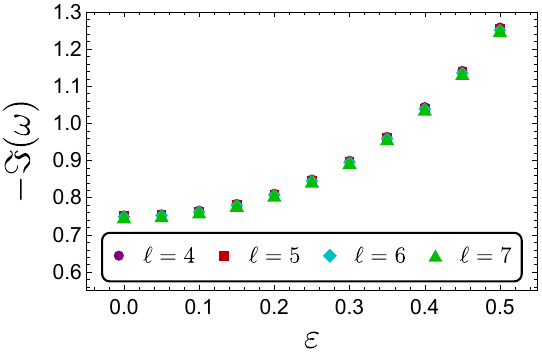}
\caption{
Real/imaginary part of Electromagnetic quasinormal frequencies, $\omega_R$/$\omega_{I}$, against the scale-dependent parameter $\varepsilon$ for the scalar effective potential $V_V(r)$.
{\bf{Top Left Panel:}} $\text{Re}(\omega)$ vs $\varepsilon$ for $M_0=1$, $G_0=1$ and $\Lambda_0=0.01$, varying $\ell$ from 4 to 7  for the fundamental mode $(n=0)$.
{\bf{Top Right Panel:}} $-\text{Im}(\omega)$ vs $\varepsilon$ for $M_0=1$, $G_0=1$ and $\Lambda_0=0.01$, varying $\ell$ from 4 to 7  for the fundamental mode $(n=0)$.
{\bf{Down Left Panel:}} $\text{Re}(\omega)$ vs $\varepsilon$ for $M_0=1$, $G_0=1$ and $\Lambda_0=0.01$, varying $\ell$ from 4 to 7 for the first exited mode $(n=1)$.
{\bf{Down Right Panel:}} $-\text{Im}(\omega)$ vs $\varepsilon$ for $M_0=1$, $G_0=1$ and $\Lambda_0=0.01$, varying $\ell$ from 4 to 7 for the first exited mode $(n=1)$.
}
\label{Fig:8}
\end{center}
\end{figure}

\begin{table}[ph!]
\centering
\caption{
Quasinormal frequencies for Electromagnetic perturbations perturbations (varying $\ell$, $n$ and $\varepsilon$) fixing $M_0=1, G_0=1, \Lambda_0=0.01$ using the WKB approximation for the effective potential $V_S(r)$.
}
{
\resizebox{0.85\columnwidth}{!}
{\begin{tabular}{c|c|cccc} 
\toprule
$\varepsilon$  &$n$  &  $\ell=4$ & $\ell=5$ & $\ell=6$ & $\ell=7$ 
\\ 
\colrule
\hline
0.00 & 0 & 1.689924\, -0.244379 i & 2.052208\, -0.245592 i & 2.411289\, -0.246320 i & 2.768382\, -0.246792 i \\
0.00 & 1 & 1.637366\, -0.741013 i & 2.008571\, -0.742212 i & 2.373964\, -0.742939 i & 2.735765\, -0.743413 i \\
0.00 & 2 & 1.535759\, -1.261978 i & 1.923246\, -1.255525 i & 2.300503\, -1.251716 i & 2.671310\, -1.249285 i \\
0.00 & 3 & 1.393329\, -1.824063 i & 1.800849\, -1.797198 i & 2.193699\, -1.781170 i & 2.576816\, -1.770876 i \\
\colrule
\hline
0.05 & 0 & 1.694178\, -0.245479 i & 2.057306\, -0.246669 i & 2.417230\, -0.247383 i & 2.775167\, -0.247846 i \\
0.05 & 1 & 1.641470\, -0.744337 i & 2.013532\, -0.745460 i & 2.379782\, -0.746141 i & 2.742439\, -0.746586 i \\
0.05 & 2 & 1.539568\, -1.267608 i & 1.927935\, -1.261001 i & 2.306075\, -1.257100 i & 2.677761\, -1.254609 i \\
0.05 & 3 & 1.396722\, -1.832156 i & 1.805138\, -1.805013 i & 2.198904\, -1.788815 i & 2.582933\, -1.778409 i \\
\colrule
\hline
0.10 & 0 & 1.707482\, -0.248800 i & 2.073469\, -0.249985 i & 2.436224\, -0.250696 i & 2.796975\, -0.251157 i \\
0.10 & 1 & 1.654145\, -0.754393 i & 2.029163\, -0.755472 i & 2.398317\, -0.756127 i & 2.763843\, -0.756554 i \\
0.10 & 2 & 1.550992\, -1.284721 i & 1.942503\, -1.277924 i & 2.323689\, -1.273911 i & 2.698356\, -1.271350 i \\
0.10 & 3 & 1.406350\, -1.856954 i & 1.818143\, -1.829261 i & 2.215149\, -1.812744 i & 2.602317\, -1.802138 i \\
\colrule
\hline
0.15 & 0 & 1.731590\, -0.254709 i & 2.102836\, -0.255909 i & 2.470792\, -0.256629 i & 2.836707\, -0.257096 i \\
0.15 & 1 & 1.677017\, -0.772306 i & 2.057498\, -0.773371 i & 2.431998\, -0.774018 i & 2.802800\, -0.774442 i \\
0.15 & 2 & 1.571415\, -1.315277 i & 1.968778\, -1.308226 i & 2.355598\, -1.304067 i & 2.735757\, -1.301415 i \\
0.15 & 3 & 1.423269\, -1.901380 i & 1.841402\, -1.872775 i & 2.244431\, -1.855734 i & 2.637401\, -1.844802 i \\
\colrule
\hline
0.20 & 0 & 1.768353\, -0.263616 i & 2.147671\, -0.264853 i & 2.523604\, -0.265597 i & 2.897437\, -0.266078 i \\
0.20 & 1 & 1.711781\, -0.799335 i & 2.100669\, -0.800418 i & 2.483384\, -0.801077 i & 2.862282\, -0.801509 i \\
0.20 & 2 & 1.602239\, -1.361485 i & 2.008642\, -1.354083 i & 2.404140\, -1.349724 i & 2.792747\, -1.346947 i \\
0.20 & 3 & 1.448490\, -1.968762 i & 1.876450\, -1.938763 i & 2.288781\, -1.920921 i & 2.690691\, -1.909491 i \\
\colrule
\hline
0.25 & 0 & 1.819631\, -0.275968 i & 2.210253\, -0.277271 i & 2.597352\, -0.278054 i & 2.982266\, -0.278562 i \\
0.25 & 1 & 1.760113\, -0.836867 i & 2.160800\, -0.838001 i & 2.555033\, -0.838690 i & 2.945275\, -0.839143 i \\
0.25 & 2 & 1.644791\, -1.425795 i & 2.063922\, -1.417903 i & 2.471617\, -1.413263 i & 2.872083\, -1.410311 i \\
0.25 & 3 & 1.482884\, -2.062824 i & 1.924709\, -2.030800 i & 2.350135\, -2.011791 i & 2.764618\, -1.999631 i \\
\colrule
\hline
0.30 & 0 & 1.887194\, -0.292249 i & 2.292757\, -0.293651 i & 2.694610\, -0.294494 i & 3.094161\, -0.295040 i \\
0.30 & 1 & 1.823564\, -0.886399 i & 2.239884\, -0.887615 i & 2.649363\, -0.888356 i & 3.054610\, -0.888842 i \\
0.30 & 2 & 1.700220\, -1.510869 i & 2.136268\, -1.502298 i & 2.560146\, -1.497266 i & 2.976330\, -1.494070 i \\
0.30 & 3 & 1.527075\, -2.187658 i & 1.987354\, -2.152796 i & 2.430188\, -2.132141 i & 2.861366\, -2.118949 i \\
\colrule
\hline
0.35 & 0 & 1.972614\, -0.312966 i & 2.397126\, -0.314506 i & 2.817684\, -0.315432 i & 3.235790\, -0.316033 i \\
0.35 & 1 & 1.903460\, -0.949509 i & 2.339659\, -0.950843 i & 2.768504\, -0.951656 i & 3.192799\, -0.952190 i \\
0.35 & 2 & 1.769388\, -1.619548 i & 2.227018\, -1.610046 i & 2.671512\, -1.604475 i & 3.107695\, -1.600941 i \\
0.35 & 3 & 1.581337\, -2.347696 i & 2.065191\, -2.308953 i & 2.530244\, -2.286037 i & 2.982707\, -2.271423 i \\
\colrule
\hline
0.40 & 0 & 2.077188\, -0.338640 i & 2.524976\, -0.340366 i & 2.968503\, -0.341405 i & 3.409386\, -0.342078 i \\
0.40 & 1 & 2.000814\, -1.027833 i & 2.461505\, -1.029324 i & 2.914182\, -1.030233 i & 3.361899\, -1.030831 i \\
0.40 & 2 & 1.852790\, -1.754812 i & 2.337106\, -1.744053 i & 2.807048\, -1.737751 i & 3.267890\, -1.733758 i \\
0.40 & 3 & 1.645517\, -2.547655 i & 2.158560\, -2.503717 i & 2.651095\, -2.477763 i & 3.129864\, -2.461228 i \\
\colrule
\hline
0.45 & 0 & 2.201886\, -0.369800 i & 2.677535\, -0.371770 i & 3.148545\, -0.372955 i & 3.616674\, -0.373725 i \\
0.45 & 1 & 2.116279\, -1.123037 i & 2.606383\, -1.124729 i & 3.087645\, -1.125759 i & 3.563433\, -1.126438 i \\
0.45 & 2 & 1.950495\, -1.919744 i & 2.466992\, -1.907317 i & 2.967569\, -1.900037 i & 3.458050\, -1.895428 i \\
0.45 & 3 & 1.718979\, -2.792533 i & 2.267258\, -2.741759 i & 2.792956\, -2.711775 i & 3.303429\, -2.692689 i \\
\colrule
\hline
0.50 & 0 & 2.347339\, -0.406970 i & 2.855630\, -0.409253 i & 3.358823\, -0.410628 i & 3.858848\, -0.411521 i \\
0.50 & 1 & 2.250133\, -1.236798 i & 2.774827\, -1.238741 i & 3.289655\, -1.239922 i & 3.798374\, -1.240700 i \\
0.50 & 2 & 2.062151\, -2.117505 i & 2.616662\, -2.102892 i & 3.153350\, -2.094326 i & 3.678717\, -2.088906 i \\
0.50 & 3 & 1.800634\, -3.087541 i & 2.390587\, -3.027888 i & 2.955459\, -2.992666 i & 3.503344\, -2.970253 i \\
\botrule
\hline
\end{tabular} 
\label{table:Second set}
}
}
\end{table}

\begin{table}[ph!]
\centering
\caption{
Quasinormal frequencies for Electromagnetic perturbations perturbations (varying $\ell$, $n$ and $\varepsilon$) fixing $M_0=1, G_0=1, \Lambda_0=0.01$ using the WKB approximation for the effective potential $V_V(r)$.
}
{
\resizebox{0.85\columnwidth}{!}
{\begin{tabular}{c|c|cccc} 
\toprule
$\varepsilon$  &$n$  &  $\ell=4$ & $\ell=5$ & $\ell=6$ & $\ell=7$ 
\\ 
\colrule
\hline
0.00 & 0 & 1.742674\, -0.248102 i & 2.096082\, -0.248175 i & 2.448854\, -0.248218 i & 2.801229\, -0.248245 i \\
0.00 & 1 & 1.691319\, -0.752110 i & 2.053118\, -0.749927 i & 2.411944\, -0.748613 i & 2.768887\, -0.747761 i \\
0.00 & 2 & 1.592276\, -1.280097 i & 1.969205\, -1.268219 i & 2.339345\, -1.261091 i & 2.704997\, -1.256487 i \\
0.00 & 3 & 1.453941\, -1.848275 i & 1.849041\, -1.814493 i & 2.233893\, -1.794069 i & 2.611382\, -1.780837 i \\
\botrule
\hline
0.05 & 0 & 1.745983\, -0.249153 i & 2.100391\, -0.249219 i & 2.454118\, -0.249257 i & 2.807421\, -0.249281 i \\
0.05 & 1 & 1.694434\, -0.755295 i & 2.057265\, -0.753078 i & 2.417069\, -0.751744 i & 2.774957\, -0.750880 i \\
0.05 & 2 & 1.595017\, -1.285525 i & 1.973032\, -1.273548 i & 2.344193\, -1.266364 i & 2.710823\, -1.261725 i \\
0.05 & 3 & 1.456157\, -1.856140 i & 1.852403\, -1.822130 i & 2.238331\, -1.801573 i & 2.616846\, -1.788259 i \\
\botrule
\hline
0.10 & 0 & 1.758778\, -0.252470 i & 2.116128\, -0.252532 i & 2.472746\, -0.252568 i & 2.828908\, -0.252591 i \\
0.10 & 1 & 1.706565\, -0.765351 i & 2.072450\, -0.763088 i & 2.435225\, -0.761728 i & 2.796032\, -0.760847 i \\
0.10 & 2 & 1.605839\, -1.302670 i & 1.987116\, -1.290486 i & 2.361404\, -1.283183 i & 2.731071\, -1.278469 i \\
0.10 & 3 & 1.465124\, -1.881037 i & 1.864880\, -1.846429 i & 2.254141\, -1.825530 i & 2.635860\, -1.812004 i \\
\botrule
\hline
0.15 & 0 & 1.782864\, -0.258428 i & 2.145476\, -0.258491 i & 2.507298\, -0.258527 i & 2.868627\, -0.258549 i \\
0.15 & 1 & 1.729389\, -0.783426 i & 2.100749\, -0.781099 i & 2.468879\, -0.779702 i & 2.834966\, -0.778797 i \\
0.15 & 2 & 1.626183\, -1.333536 i & 2.013331\, -1.320993 i & 2.393266\, -1.313486 i & 2.768437\, -1.308645 i \\
0.15 & 3 & 1.481962\, -1.925953 i & 1.888062\, -1.890265 i & 2.283360\, -1.868746 i & 2.670894\, -1.854836 i \\
\botrule
\hline
0.20 & 0 & 1.820147\, -0.267448 i & 2.190745\, -0.267515 i & 2.560481\, -0.267553 i & 2.929682\, -0.267577 i \\
0.20 & 1 & 1.764659\, -0.810813 i & 2.144342\, -0.808394 i & 2.520628\, -0.806942 i & 2.894767\, -0.806002 i \\
0.20 & 2 & 1.657515\, -1.380389 i & 2.053609\, -1.367287 i & 2.442163\, -1.359458 i & 2.825739\, -1.354416 i \\
0.20 & 3 & 1.507758\, -1.994291 i & 1.923540\, -1.956900 i & 2.328064\, -1.934399 i & 2.724491\, -1.919875 i \\
\botrule
\hline
0.25 & 0 & 1.872547\, -0.279991 i & 2.254262\, -0.280067 i & 2.635032\, -0.280110 i & 3.015213\, -0.280137 i \\
0.25 & 1 & 1.814111\, -0.848942 i & 2.205404\, -0.846388 i & 2.593076\, -0.844857 i & 2.978460\, -0.843868 i \\
0.25 & 2 & 1.701231\, -1.445748 i & 2.109831\, -1.431823 i & 2.510440\, -1.423517 i & 2.905773\, -1.418176 i \\
0.25 & 3 & 1.543466\, -2.089873 i & 1.972794\, -2.049978 i & 2.390237\, -2.026023 i & 2.799123\, -2.010586 i \\
\botrule
\hline
0.30 & 0 & 1.941893\, -0.296557 i & 2.338254\, -0.296646 i & 2.733567\, -0.296696 i & 3.128227\, -0.296728 i \\
0.30 & 1 & 1.879362\, -0.899358 i & 2.285982\, -0.896614 i & 2.688685\, -0.894972 i & 3.088914\, -0.893911 i \\
0.30 & 2 & 1.758551\, -1.532363 i & 2.183702\, -1.517272 i & 2.600260\, -1.508289 i & 3.011144\, -1.502520 i \\
0.30 & 3 & 1.589801\, -2.216919 i & 2.037065\, -2.173495 i & 2.471626\, -2.147481 i & 2.897016\, -2.130745 i \\
\botrule
\hline
0.35 & 0 & 2.029824\, -0.317673 i & 2.444719\, -0.317780 i & 2.858440\, -0.317841 i & 3.271431\, -0.317879 i \\
0.35 & 1 & 1.961802\, -0.963703 i & 2.387867\, -0.960696 i & 2.809631\, -0.958900 i & 3.228681\, -0.957740 i \\
0.35 & 2 & 1.830415\, -1.643186 i & 2.276622\, -1.626489 i & 2.713458\, -1.616570 i & 3.144101\, -1.610208 i \\
0.35 & 3 & 1.647143\, -2.380015 i & 2.117231\, -2.331761 i & 2.573588\, -2.302915 i & 3.019988\, -2.284387 i \\
\botrule
\hline
0.40 & 0 & 2.137699\, -0.343885 i & 2.575326\, -0.344016 i & 3.011627\, -0.344091 i & 3.447102\, -0.344138 i \\
0.40 & 1 & 2.062513\, -1.043689 i & 2.512493\, -1.040326 i & 2.957687\, -1.038321 i & 3.399862\, -1.037027 i \\
0.40 & 2 & 1.917397\, -1.781333 i & 2.389583\, -1.762470 i & 2.851417\, -1.751287 i & 3.306399\, -1.744123 i \\
0.40 & 3 & 1.715448\, -2.584086 i & 2.213702\, -2.529359 i & 2.696972\, -2.496708 i & 3.169307\, -2.475764 i \\
\botrule
\hline
0.45 & 0 & 2.266552\, -0.375748 i & 2.731359\, -0.375911 i & 3.194652\, -0.376005 i & 3.657005\, -0.376064 i \\
0.45 & 1 & 2.182215\, -1.141075 i & 2.660881\, -1.137239 i & 3.134153\, -1.134955 i & 3.604021\, -1.133483 i \\
0.45 & 2 & 2.019658\, -1.950053 i & 2.523110\, -1.928326 i & 3.015002\, -1.915464 i & 3.499217\, -1.907237 i \\
0.45 & 3 & 1.794227\, -2.834352 i & 2.326368\, -2.771114 i & 2.842054\, -2.733433 i & 3.345611\, -2.709291 i \\
\botrule
\hline
0.50 & 0 & 2.417082\, -0.413823 i & 2.913702\, -0.414028 i & 3.408581\, -0.414146 i & 3.902379\, -0.414221 i \\
0.50 & 1 & 2.321256\, -1.257647 i & 2.833619\, -1.253193 i & 3.339837\, -1.250544 i & 3.842176\, -1.248839 i \\
0.50 & 2 & 2.136932\, -2.152715 i & 2.677248\, -2.127250 i & 3.204540\, -2.112195 i & 3.723143\, -2.102575 i \\
0.50 & 3 & 1.882529\, -3.136359 i & 2.454594\, -3.062068 i & 3.008522\, -3.017834 i & 3.548897\, -2.989520 i \\
\botrule
\hline
\end{tabular} 
\label{table:Third set}
}
}
\end{table}

In this case, we have computed the electromagnetic perturbations for both the scalar and vector case according to the effective potentials \eqref{VSelecto}-\eqref{VVelecto} respectively. As in the previous case, we have collected our numerical result in tables \eqref{table:Second set} - \eqref{table:Third set} and different figures (see Fig.~\eqref{Fig:6}, \eqref{Fig:7} and \eqref{Fig:8}).
In particular, we show, in Figure \eqref{Fig:6}, the behavior of the fundamental quasinormal modes (Top-left panel) and the first excited modes (Top-right panel) of the effective potential $V_S(r)$ for different values of the scale-dependent parameter $\varepsilon$. 
In the same figure, second row, we show the same quantities, but now using the potential $V_V(r)$.
Similarly as the scalar case, in Fig.~\eqref{Fig:7} and  Fig.~\eqref{Fig:8}, we show the direct impact of a change in the scale-dependent parameter $\varepsilon$ for the real and imaginary parts of the fundamental mode ($n=0$) in the first row and the first excited mode ($n=1$) in the second row, taking $V_S(r)$ in Fig.~\eqref{Fig:7} and $V_V(r)$ in Fig.~\eqref{Fig:8}. 
Thus, our results confirm that this black hole background is stable against electromagnetic (scalar and vector) perturbations. Once, this is because the imaginary part of the frequencies is negative in all cases studied. 
Moreover, both the real part of the frequency and the magnitude of its imaginary part increase with increasing $\varepsilon$, pointing to a stronger correction effect.
%

\subsection{Dirac Perturbations}

\begin{figure*}
    \centering
    \includegraphics[width=0.49\textwidth]{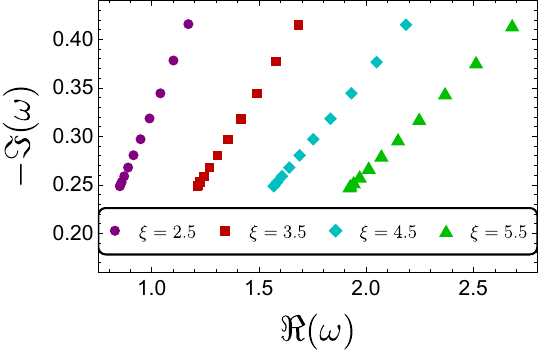}
    \
    \includegraphics[width=0.49\textwidth]{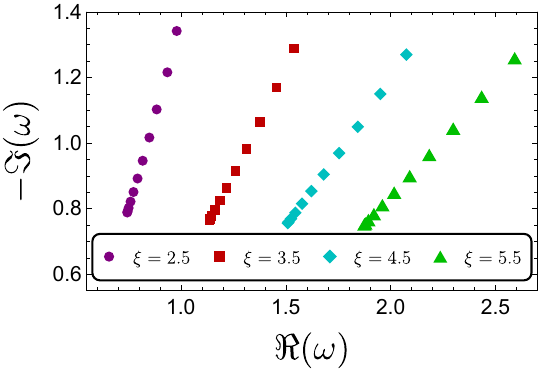}
    \caption{
    Quasinormal modes of massless Dirac perturbations for different values of the scale-dependent parameter $\varepsilon$ and the angular number $\ell$, keeping fixed $\{ M_0 =  1, G_0 = 1, \Lambda_0  = +0.01\}$.
    {\bf{Left panel:}} Quasinormal modes for the fundamental mode $(n=0)$. 
    {\bf{Right panel:}} Quasinormal modes for the first exited mode $(n=1)$. 
    }
    \label{Fig:9}
\end{figure*}

\begin{figure} 
\begin{center}
\includegraphics[width=0.49\linewidth]{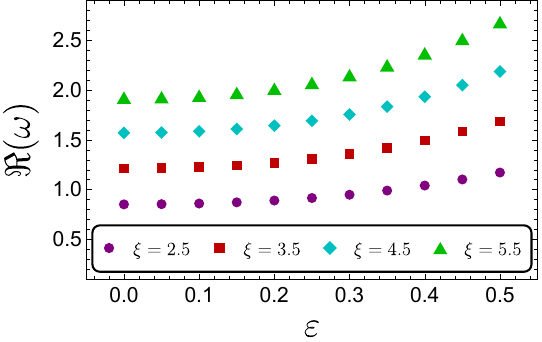} 
\includegraphics[width=0.49\linewidth]{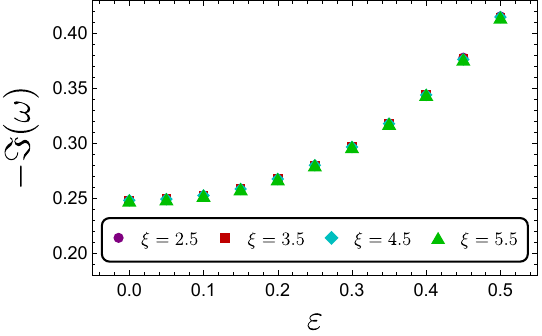}
\\
\vspace{0.5cm}
\includegraphics[width=0.49\linewidth]{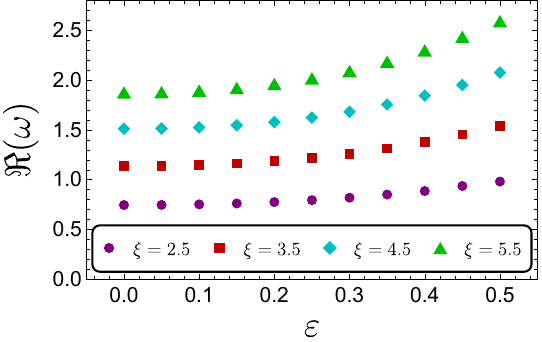} 
\includegraphics[width=0.49\linewidth]{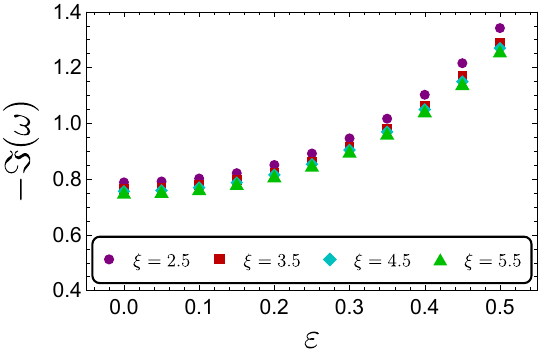}
\caption{
Real/imaginary part of massless Dirac quasinormal frequencies, $\omega_R$/$\omega_{I}$, against the dimensionless scale-dependent parameter $\varepsilon$.
{\bf{Top Left Panel:}} $\text{Re}(\omega)$ vs $\varepsilon$ for $M_0=1$, $G_0=1$, varying $\xi$ from 5/2 to 11/2  for the fundamental mode $(n=0)$.
{\bf{Top Right Panel:}} $-\text{Im}(\omega)$ vs $\varepsilon$ for $M_0=1$, $G_0=1$, varying $\xi$ from 5/2 to 11/2  for the fundamental mode $(n=0)$.
{\bf{Down Left Panel:}} $\text{Re}(\omega)$ vs $\varepsilon$ for $M_0=1$, $G_0=1$, varying $\xi$ from 5/2 to 11/2 for the first exited mode $(n=1)$.
{\bf{Down Right Panel:}} $-\text{Im}(\omega)$ vs $\varepsilon$ for $M_0=1$, $G_0=1$, varying $\xi$ from 5/2 to 11/2 for the first exited mode $(n=1)$.
}
\label{Fig:10}
\end{center}
\end{figure}

Finally, Figure~\eqref{Fig:9} illustrates the behavior of (i) the fundamental quasinormal modes (left panel) and (ii) the first excited modes (right panel) for $0.0 \leq \varepsilon \leq 0.5$ and four different values of $\xi$, using the effective potential $V_+(r)$ previously defined for massless Dirac fields.
In this case, we observe a trend consistent with the two cases studied earlier: for smaller values of $\xi$ (equivalently, smaller $\ell$), the quasinormal frequencies are only weakly affected by changes in $\varepsilon$. 
As shown also in Fig.~\eqref{Fig:10}, however, the deviation becomes significantly more pronounced as $\xi$ increases. This indicates that the higher the value of $\xi$, the greater the sensitivity of the quasinormal frequencies to variations in the parameter $\varepsilon$.
As can be seen in Table~$\eqref{table:Fourth set}$, all the numerical values have a negative imaginary part, which shows that the background is stable against massless Dirac perturbations.

\begin{table}[ph!]
\centering
\caption{
Quasinormal frequencies for massless Dirac perturbations (varying $\ell$, $n$ and $\varepsilon$) fixing $M_0=1, G_0=1, \Lambda_0=0.01$ using the WKB approximation.
}
{
\resizebox{0.85\columnwidth}{!}
{\begin{tabular}{c|c|cccc} 
\toprule
$\varepsilon$  &$n$  &  $\xi=5/2$ & $\xi=7/2$ & $\xi=9/2$ & $\xi=11/2$ 
\\ 
\colrule
\hline
0.00 & 0 & 0.851797\, -0.248403 i & 1.213758\, -0.248364 i & 1.569461\, -0.248402 i & 1.923042\, -0.248392 i \\
0.00 & 1 & 0.744882\, -0.788016 i & 1.136224\, -0.764281 i & 1.510101\, -0.755891 i & 1.874985\, -0.752028 i \\
0.00 & 2 & 0.596939\, -1.372009 i & 1.010890\, -1.313130 i & 1.406442\, -1.286075 i & 1.787475\, -1.271921 i \\
0.00 & 3 & 0.411975\, -1.971223 i & 0.853360\, -1.882702 i & 1.272028\, -1.836133 i & 1.670592\, -1.809060 i \\
\botrule
\hline
0.05 & 0 & 0.853777\, -0.249426 i & 1.216645\, -0.249388 i & 1.573219\, -0.249426 i & 1.927656\, -0.249417 i \\
0.05 & 1 & 0.746438\, -0.791298 i & 1.138792\, -0.767441 i & 1.513620\, -0.759010 i & 1.879411\, -0.755130 i \\
0.05 & 2 & 0.597924\, -1.377645 i & 1.012962\, -1.318525 i & 1.409555\, -1.291360 i & 1.791564\, -1.277151 i \\
0.05 & 3 & 0.412157\, -1.979237 i & 0.854761\, -1.890365 i & 1.274595\, -1.843617 i & 1.674222\, -1.816452 i \\
\botrule
\hline
0.10 & 0 & 0.860254\, -0.252729 i & 1.226089\, -0.252684 i & 1.585512\, -0.252727 i & 1.942757\, -0.252720 i \\
0.10 & 1 & 0.751522\, -0.801926 i & 1.147169\, -0.777638 i & 1.525113\, -0.769069 i & 1.893876\, -0.765132 i \\
0.10 & 2 & 0.601181\, -1.395941 i & 1.019692\, -1.335976 i & 1.419693\, -1.308426 i & 1.804898\, -1.294032 i \\
0.10 & 3 & 0.412873\, -2.005266 i & 0.859291\, -1.915188 i & 1.282919\, -1.867822 i & 1.686027\, -1.840338 i \\
\botrule
\hline
0.15 & 0 & 0.872076\, -0.258679 i & 1.243344\, -0.258628 i & 1.607980\, -0.258678 i & 1.970358\, -0.258674 i \\
0.15 & 1 & 0.760704\, -0.821143 i & 1.162417\, -0.796069 i & 1.546066\, -0.787226 i & 1.920273\, -0.783179 i \\
0.15 & 2 & 0.606918\, -1.429115 i & 1.031869\, -1.367597 i & 1.438093\, -1.339289 i & 1.829158\, -1.324531 i \\
0.15 & 3 & 0.413869\, -2.052542 i & 0.867421\, -1.960248 i & 1.297931\, -1.911668 i & 1.707413\, -1.883555 i \\
\botrule
\hline
0.20 & 0 & 0.890131\, -0.267708 i & 1.269730\, -0.267637 i & 1.642353\, -0.267702 i & 2.012594\, -0.267701 i \\
0.20 & 1 & 0.774616\, -0.850436 i & 1.185602\, -0.824061 i & 1.578031\, -0.814800 i & 1.960588\, -0.810564 i \\
0.20 & 2 & 0.615519\, -1.479808 i & 1.050175\, -1.415737 i & 1.466026\, -1.386255 i & 1.866076\, -1.370878 i \\
0.20 & 3 & 0.415234\, -2.124858 i & 0.879349\, -2.028971 i & 1.320574\, -1.978514 i & 1.739795\, -1.949331 i \\
\botrule
\hline
0.25 & 0 & 0.915313\, -0.280328 i & 1.306574\, -0.280179 i & 1.690377\, -0.280259 i & 2.071620\, -0.280266 i \\
0.25 & 1 & 0.794000\, -0.891609 i & 1.217791\, -0.863151 i & 1.622525\, -0.853227 i & 2.016803\, -0.848725 i \\
0.25 & 2 & 0.627790\, -1.551173 i & 1.075371\, -1.483152 i & 1.504637\, -1.451833 i & 1.917352\, -1.435576 i \\
0.25 & 3 & 0.417920\, -2.226635 i & 0.895515\, -2.125373 i & 1.351518\, -2.072004 i & 1.784537\, -2.041301 i \\
\botrule
\hline
0.30 & 0 & 0.948377\, -0.296873 i & 1.355132\, -0.296767 i & 1.753716\, -0.296838 i & 2.149495\, -0.296860 i \\
0.30 & 1 & 0.818539\, -0.945787 i & 1.259952\, -0.915042 i & 1.680922\, -0.904023 i & 2.090747\, -0.899157 i \\
0.30 & 2 & 0.641378\, -1.645436 i & 1.108122\, -1.572904 i & 1.554785\, -1.538650 i & 1.984375\, -1.521159 i \\
0.30 & 3 & 0.416727\, -2.361596 i & 0.916282\, -2.253902 i & 1.390910\, -2.195953 i & 1.842386\, -2.163087 i \\
\botrule
\hline
0.35 & 0 & 0.990142\, -0.318283 i & 1.416499\, -0.317852 i & 1.833861\, -0.317978 i & 2.248075\, -0.318031 i \\
0.35 & 1 & 0.850301\, -1.016450 i & 1.312538\, -0.981118 i & 1.754441\, -0.968927 i & 2.184083\, -0.963621 i \\
0.35 & 2 & 0.661690\, -1.768182 i & 1.147567\, -1.687394 i & 1.617348\, -1.649835 i & 2.068615\, -1.630828 i \\
0.35 & 3 & 0.421806\, -2.536586 i & 0.938825\, -2.418184 i & 1.439309\, -2.354959 i & 1.914771\, -2.319464 i \\
\botrule
\hline
0.40 & 0 & 1.040829\, -0.344183 i & 1.491614\, -0.344139 i & 1.932038\, -0.344143 i & 2.368918\, -0.344348 i \\
0.40 & 1 & 0.885228\, -1.102214 i & 1.376613\, -1.064037 i & 1.843705\, -1.049229 i & 2.298144\, -1.043953 i \\
0.40 & 2 & 0.675687\, -1.918339 i & 1.196011\, -1.831593 i & 1.691501\, -1.787384 i & 2.171176\, -1.767914 i \\
0.40 & 3 & 0.407479\, -2.752788 i & 0.967720\, -2.625174 i & 1.493476\, -2.551850 i & 2.002707\, -2.515366 i \\
\botrule
\hline
0.45 & 0 & 1.101731\, -0.378026 i & 1.581308\, -0.376894 i & 2.049272\, -0.376351 i & 2.513213\, -0.376306 i \\
0.45 & 1 & 0.936021\, -1.215404 i & 1.455165\, -1.169189 i & 1.950767\, -1.149352 i & 2.433637\, -1.141532 i \\
0.45 & 2 & 0.720957\, -2.113009 i & 1.263668\, -2.015521 i & 1.783993\, -1.960850 i & 2.291586\, -1.934464 i \\
0.45 & 3 & 0.446089\, -3.026542 i & 1.024829\, -2.887853 i & 1.569460\, -2.801015 i & 2.103626\, -2.753441 i \\
\botrule
\hline
0.50 & 0 & 1.171323\, -0.415522 i & 1.685451\, -0.414414 i & 2.186144\, -0.414857 i & 2.681794\, -0.414512 i \\
0.50 & 1 & 0.980630\, -1.341283 i & 1.539186\, -1.287496 i & 2.074865\, -1.269385 i & 2.591147\, -1.258407 i \\
0.50 & 2 & 0.732412\, -2.333948 i & 1.316036\, -2.221625 i & 1.889991\, -2.169130 i & 2.430333\, -2.134375 i \\
0.50 & 3 & 0.412332\, -3.345315 i & 1.035037\, -3.185536 i & 1.654827\, -3.100273 i & 2.218323\, -3.039610 i \\
\botrule
\hline
\end{tabular} 
\label{table:Fourth set}
}
}
\end{table}

\section{Concluding remarks}
\label{sec:conclu}

In this paper, we present a refined derivation of a five-dimensional scale-dependent black hole solution with a positive cosmological constant. We examine the black hole's response to i) scalar, ii) electromagnetic, and iii) Dirac perturbations to evaluate the impact of quantum corrections on the classical background. These corrections are parameterized by the dimensionless quantity ($\varepsilon$). Given the small magnitude of quantum effects, we utilize a first-order approximation for the lapse function to calculate the quasinormal (QN) frequencies, an approach valid for ($\varepsilon \ll 1$). Our findings, summarized in detailed tables and figures, demonstrate that the black hole is stable across all perturbation types, as indicated by the consistently negative imaginary parts of the QN frequencies.
Finally, we investigate whether such a small quantum correction can leave detectable imprints on the quasinormal spectrum. Although the parameter $\varepsilon$ is small by definition, the modified background effectively mimics the presence of a small charge, due to the scale dependence of both the Newtonian and cosmological couplings.


\section*{Acknowledgments}

The authors would like to express their gratitude for the comments and suggestions made by the reviewer, which helped significantly improve the manuscript.
A.~R. would like to express his gratitude to Silesian University in Opava, Czech Republic, for their financial support.
The creation of this article was supported by the grant program Vouchers for Universities in the Moravian-Silesian Region (registration number CZ.10.03.01/00/23\_042/0000390).
This article is based upon work from COST Action FuSe, CA24101, supported by COST (European Cooperation in Science and Technology).

\section*{Data Availability Statement}
\noindent

\verb|The data generated in this paper is presented in tables and figures within the|
\verb|main body of the text.|

\section*{Code Availability Statement}
\noindent

\verb|Code will be made available on reasonable request.|

\bibliographystyle{utphys}
\bibliography{Library}

\end{document}